\documentclass{article}

\usepackage{hyperref}

\usepackage{amsmath}
\usepackage{amssymb}

\usepackage{algpseudocode}
\usepackage{algorithm}

\usepackage{natbib}

\usepackage{booktabs}

\usepackage{graphicx}

\usepackage{subcaption}

\newcommand{\argmax}{\operatornamewithlimits{argmax}}
\newcommand{\argmin}{\operatornamewithlimits{argmin}}
\newcommand{\pmf}{\operatorname{pmf}}

\newcommand{\upp}[1]{\ensuremath{{\| u^{(m)} \|_{{#1} p^*}^{{#1} p^*}}}}
\newcommand{\vpp}[1]{\ensuremath{{\| v^{(m)} \|_{{#1} p^*}^{{#1} p^*}}}}
\newcommand{\vp}[1]{\ensuremath{{\| v^{(m)} \|_{{#1}}^{{#1}}}}}

\usepackage{mathtools}

\DeclarePairedDelimiter{\ceil}{\lceil}{\rceil}
\DeclarePairedDelimiter{\floor}{\lfloor}{\rfloor}

\title{A Bounded $p$-norm Approximation of Max-Convolution for
  Sub-Quadratic Bayesian Inference on Additive Factors}

\author{Julianus Pfeuffer\\
       Eberhard Karls Universit\"at T\"ubingen\\
       \and
       Oliver Serang\\
       Freie Universit\"at Berlin\\
       Department of Informatik\\
       Takustr. 9, 14195 Berlin, Germany / \\
       The Leibniz-Institute of Freshwater Ecology and Inland Fisheries (IGB)\\
         M\"uggelsee 310, 12587 Berlin, Germany\\
         orserang@uw.edu 
       }

\date{\today}

\begin{document}

\sloppy

\maketitle

\begin{abstract}%
Max-convolution is an important problem closely resembling standard
convolution; as such, max-convolution occurs frequently across many
fields. Here we extend the method with fastest known worst-case
runtime, which can be applied to nonnegative vectors by numerically
approximating the Chebyshev norm $\| \cdot \|_\infty$, and use this
approach to derive two numerically stable methods based on the idea of
computing $p$-norms via fast convolution: The first method proposed,
with runtime in $O( k \log(k) \log(\log(k)) )$ (which is less than $18
k \log(k)$ for any vectors that can be practically realized), uses the
$p$-norm as a direct approximation of the Chebyshev norm. The second
approach proposed, with runtime in $O( k \log(k) )$ (although in
practice both perform similarly), uses a novel null space projection
method, which extracts information from a sequence of $p$-norms to
estimate the maximum value in the vector (this is equivalent to
querying a small number of moments from a distribution of bounded
support in order to estimate the maximum). The $p$-norm approaches are
compared to one another and are shown to compute an approximation of
the Viterbi path in a hidden Markov model where the transition matrix
is a Toeplitz matrix; the runtime of approximating the Viterbi path is
thus reduced from $O( n k^2 )$ steps to $O( n k \log(k))$ steps in
practice, and is demonstrated by inferring the U.S. unemployment rate
from the S\&P 500 stock index.
\end{abstract}

\section{Introduction}


Max-convolution occurs frequently in signal processing and Bayesian
inference: it is used in image analysis~\citep{Ritter2000b}, in network
calculus~\citep{Boyer2013}, in economic equilibrium
analysis~\citep{Sun2002b}, and in a probabilistic variant of
combinatoric generating functions, wherein information on a sum of
values into their most probable constituent parts (\emph{e.g.}
identifying proteins from mass
spectrometry~\citep{Serang2010a,Serang2014b}). Max-convolution operates
on the semi-ring $(max, \times)$, meaning that it behaves identically
to a standard convolution, except it employs a $\max$ operation in
lieu of the $+$ operation in standard convolution (max-convolution is
also equivalent to min-convolution, also called infimal convolution,
which operates on the tropical semi-ring $(min, +)$). Due to the
importance and ubiquity of max-convolution, substantial effort has
been invested into highly optimized implementations (\emph{e.g.},
implementations of the quadratic method on GPUs; \citealp{Zach2008}).

Max-convolution can be defined using vectors (or discrete random
variables, whose probability mass functions are analogous to
nonnegative vectors) with the relationship $M=L+R$. Given the target
sum $M=m$, the max-convolution finds the largest values $L[\ell]$ and
$R[r]$ for which $m=\ell + r$.
\begin{eqnarray*}
M[m] & = & \max_{\ell,r :\,m = \ell+r} L[\ell] R[r] \\
&=& \max_\ell L[\ell] R[{m-\ell}]\\
&=& \left( L ~*_{\max}~ R \right)[m] \\
\end{eqnarray*}
where $*_{\max}$ denotes the max-convolution operator. In
probabilistic terms, this is equivalent to finding the highest
probability of the joint events $\Pr(L=\ell, R=r)$ that would produce
each possible value of the sum $M=L+R$ (note that in the probabilistic
version, the vector $M$ would subsequently need to be normalized so
that its sum is 1).\newline

Although applications of max-convolution are numerous, only a small
number of methods exist for solving it~\citep{Serang2015}. These
methods fall into two main categories, each with their own drawbacks:
The first category consists of very accurate methods that are have
worst-case runtimes either quadratic~\citep{Bussieck1994} or slightly
more efficient than quadratic in the
worst-case~\citep{Bremner2006}. Conversely, the second type of
method computes a numerical approximation to the desired result, but
in $O(k \log_2(k))$ steps; however, no bound for the numerical accuracy
of this method has been derived~\citep{Serang2015}.

While the two approaches from the first category of methods for
solving max-convolution do so by either using complicated sorting
routines or by creating a bijection to an optimization problem, the
numerical approach solves max-convolution by showing an equivalence
between $*_{\max}$ and the process of first generating a vector
$u^{(m)}$ for each index $m$ of the result (where $u^{(m)}[\ell] =
L[\ell] R[{m-\ell}]$ for all in-bounds indices) and subsequently
computing the maximum $M[m] = \max_\ell u^{(m)}[\ell]$. When $L$ and $R$
are nonnegative, the maximization over the vector $u^{(m)}$ can be
computed exactly via the Chebyshev norm
\begin{eqnarray*}
M[m] &=& \max_\ell u^{(m)}[\ell] \\
&=& \lim_{p \to \infty} \| u^{(m)} \|_p\\
\end{eqnarray*}
but requires $O(k^2)$ steps (where $k$ is the length of vectors $L$
and $R$). However, once a fixed $p^*$-norm is chosen, the
approximation corresponding to that $p^*$ can be computed by expanding
the $p^*$-norm to yield
\begin{eqnarray*}
  \lim_{p \to \infty} \| u^{(m)} \|_p &=& \lim_{p \to \infty} {\left( \sum_\ell {\left( u^{(m)}[\ell] \right)}^{p} \right)}^{\frac{1}{p}} \\
  & \approx & {\left( \sum_\ell {\left( u^{(m)}[\ell] \right)}^{p^*} \right)}^{\frac{1}{p^*}}\\
  & = & {\left( \sum_\ell {L[\ell]}^{p^*} ~ {R[{m-\ell}]}^{p^*} \right)}^{\frac{1}{p^*}}\\
  & = & {\left( \sum_\ell {\left(L^{p^*}\right)}[\ell] ~ {\left(R^{p^*}\right)}[{m-\ell}] \right)}^{\frac{1}{p^*}}\\
  & = & {\left( L^{p^*} ~*~ R^{p^*} \right)}^{\frac{1}{p^*}}[m]
\end{eqnarray*}
where $L^{p^*} = ~ \langle~ {\left( L[0] \right)}^{p^*}, {\left(
  L[1] \right)}^{p^*}, ~\ldots,~{\left( L[{k-1}] \right)}^{p^*}
~\rangle$ and $*$ denotes standard convolution. The standard
convolution can be done via fast Fourier transform (FFT) in $O(k
\log_2(k))$ steps, which is substantially more efficient than the
$O(k^2)$ required by the naive method ({\bf
  Algorithm~\ref{algorithm:numericalMaxConvolveGivenPStar}}).

To date, the numerical method has currently demonstrated the best
speed-accuracy trade-off on Bayesian inference tasks, and can be
generalized to multiple dimensions (\emph{i.e.}, tensors). In
particular, they have been used with probabilistic convolution
trees~\citep{Serang2014b} to efficiently compute the most probable
values of discrete random variables $X_0, X_1, \ldots X_{n-1}$ for
which the sum is known $X_0 + X_1 + \ldots X_{n-1} =
y$~\citep{Serang2014b}. The one-dimensional variant of this problem
(\emph{i.e.}, where each $X_i$ is a one-dimensional vector) solves the
probabilistic generalization of the subset sum problem, while the
two-dimensional variant (\emph{i.e.}, where each $X_i$ is a
one-dimensional matrix) solves the generalization of the knapsack
problem (note that these problems are not NP-hard in this specific
case, because we assume an evenly-spaced discretization of the
possible values of the random variables).

However, despite the practical performance that has been demonstrated
by the numerical method, only cursory analysis has been performed to
formalize the influence of the value of $p^*$ on the accuracy of the
result and to bound the error of the $p^*$-norm
approximation. Optimizing the choice of $p^*$ is non-trivial: Larger
values of $p^*$ more closely resemble a true maximization under the
$p^*$-norm, but result in underflow (note that in {\bf
  Algorithm~\ref{algorithm:numericalMaxConvolveGivenPStar}}, the
maximum values of both $L$ and $R$ can be divided out and then
multiplied back in after max-convolution so that overflow is not an
issue). Conversely, smaller values of $p^*$ suffer less underflow, but
compute a norm with less resemblance to maximization. Here we perform
an in-depth analysis of the influence of $p^*$ on the accuracy of
numerical max-convolution, and from that analysis we construct a
modified piecewise algorithm, on which we demonstrate bounds on the
worst-case absolute error. This modified algorithm, which runs in $O(
k \log(k) \log(\log(k)) )$ steps, is demonstrated using a hidden
Markov model describing the relationship between U.S. unemployment and
the S\&P 500 stock index.

We then extend the modified algorithm and introduce a second modified
algorithm, which not only uses a single $p$-norm as a means of
approximating the Chebyshev norm, but instead uses a sequence of
$p$-norms and assembles them using a projection as a means to
approximate the Chebyshev norm. Using numerical simulations as
evidence, we make a conjecture regarding the relative error of the
null space projection method. In practice, this null space projection
algorithm is shown to have similar runtime and higher accuracy when
compared with the piecewise algorithm.

\section{Methods} \label{documentclasses}

We begin by outlining and comparing three numerical methods for
max-convolution. By analyzing the benefits and deficits of each of
these methods, we create improved variants. All of these methods will
make use of the basic numerical max-convolution idea summarized in the
introduction, and as such we first declare a method for computing the
numerical max-convolution estimate for a given $p^*$ as {\tt
  numericalMaxConvolveGivenPStar} ({\bf
  Algorithm~\ref{algorithm:numericalMaxConvolveGivenPStar}}).

\begin{algorithm}
  \caption{ {\bf Numerical max-convolution given a fixed $p^*$}, a
    numerical method to estimate the max-convolution of two PMFs or
    nonnegative vectors. The parameters are two nonnegative vectors
    $L'$ and $R'$ (both scaled so that they have maximal element 1)
    and the numerical value $p^*$ used for computation. The return
    value is a numerical estimate of the max-convolution $L' ~*_{\max}~
    R'$.}

  \label{algorithm:numericalMaxConvolveGivenPStar}
  \begin{small}
    \begin{algorithmic}[1]
      \Procedure{numericalMaxConvolveGivenPStar}{$L'$, $R'$, $p^*$}
      
      \State $\forall \ell, ~ vL[\ell] \gets { L[\ell] }^{p^*}$

      \State $\forall r, ~ vR[r] \gets { R[r] }^{p^*}$

      \State $vM \gets vL ~*~ vR$ \Comment{Standard FFT convolution is used here}

      \State $\forall m, ~ M'[m] \gets { vM[m] }^{\frac{1}{p^*}}$
      
      \State \Return $M'$ 
      \EndProcedure
    \end{algorithmic}
  \end{small}
\end{algorithm}

\subsection{Fixed Low-Value $p^*=8$ Method:} 
The effects of underflow will be minimal (as it is not very far from
standard FFT convolution, an operation with high numerical stability),
but it can still be imprecise due to numerical ``bleed-in''
(\emph{i.e.} error due to contributions from non-maximal terms for a
given $u^{(m)}$ because the $p^*$-norm is not identical to the
Chebyshev norm). Overall, this will perform well on indices where the
exact value of the result is small, but perform poorly when the exact
value of the result is large.

\subsection{Fixed High-Value $p^*=64$ Method:} 
As noted above, will offer the converse pros and cons compared to
using a low $p^*$: numerical artifacts due to bleed-in will be smaller
(thus achieving greater performance on indices where the exact values
of the result are larger), but underflow may be significant (and
therefore, indices where the exact results of the max-convolution are
small will be inaccurate).

\subsection{Higher-Order Piecewise Method:}
The higher-order piecewise method formalizes the empirical cutoff
values found in Serang 2015; previously, numerical stability
boundaries were found for each $p^*$ by computing both the exact
max-convolution (via the naive $O(k^2)$ method) and via the numerical
method using the ascribed value of $p^*$, and finding the value below
which the numerical values experienced a high increase in relative
absolute error.

\begin{figure*}
  \centering
  \includegraphics[width=5in]{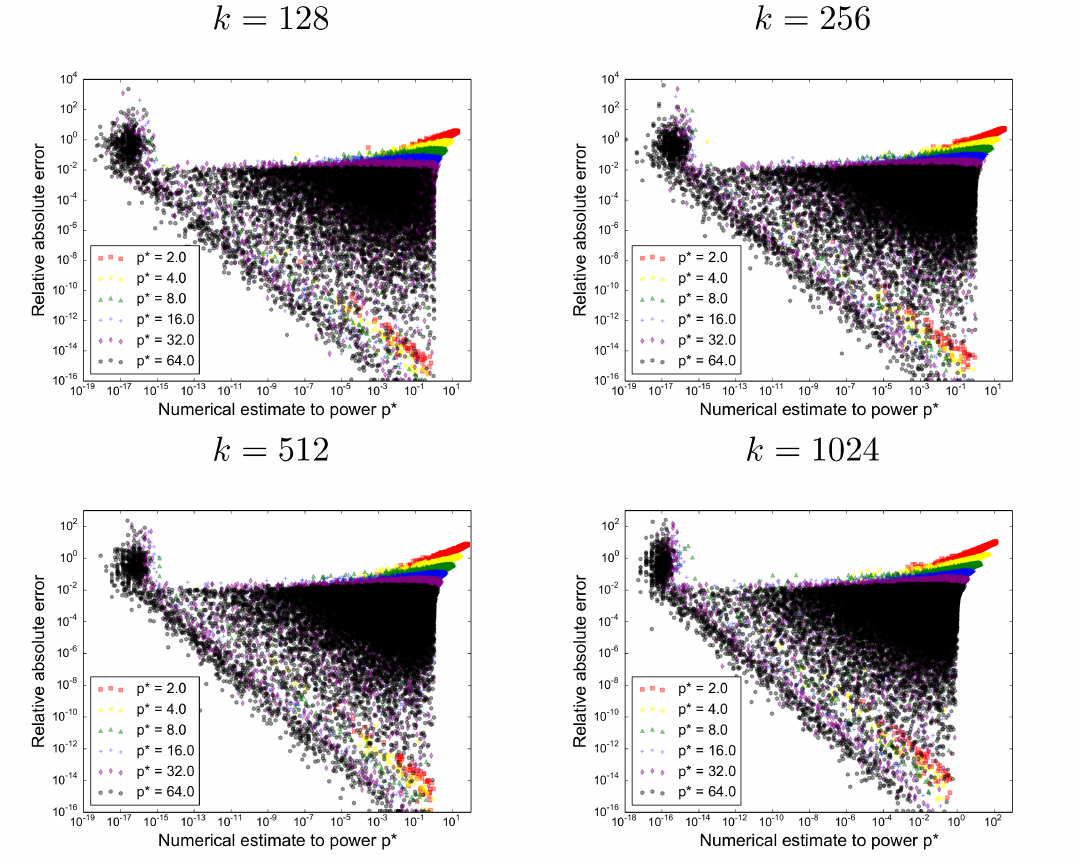}
  \caption{{\bf Empirical estimate of $\tau$ to construct a piecewise
      method.} For each $k \in \{128, 256,512,1024\}$, 32 replicate
    max-convolutions (on vectors filled with uniform values) are
    performed. Error from two sources can be seen: error due to
    underflow is depicted in the sharp left mode, whereas error due to
    imperfect approximation, where $\| \cdot \|_{p^*} > \| \cdot
    \|_\infty$ can be seen in the gradual mode on the right. Error due
    to $p^*$-norm approximation is significantly smaller when $p^*$ is
    larger (thereby flattening the right mode), but larger $p^*$
    values are more susceptible to underflow, pushing more indices
    into the left mode. Regardless of the value of $k$, error due to
    underflow occurs when ${\left( \| \cdot \|_{p^*}\right)}^{p^*}$
    goes below $\approx {10}^{-15}$; this is approximately the
    numerical tolerance for $\tau$ described by the {\tt numpy}
    documentation. Therefore, at each index $m$ we can construct a
    piecewise method that uses the largest value of $p^*$ for which
    the FFT convolution result is not close to the machine precision
    (\emph{i.e.}, $({\| u^{(m)} \|}_{p^*})^{p^*} \geq \tau$ for some
    $\tau > {10}^{-15}$).
  \label{figure:badErrorsForAllPStar}}
\end{figure*}

Those previously observed empirical numerical stability boundaries can
be formalized by using the fact that the employed {\tt numpy}
implementation of FFT convolution has high accuracy on indices where
the result has a value $\geq \tau$ relative to the maximum value;
therefore, if the arguments $L$ and $R$ are both normalized so that
each has a maximum value of 1, the fast max-convolution approximation
is numerically stable for any index $m$ where the result of the FFT
convolution, \emph{i.e.}  $vM[m]$, is $\geq \tau$. The {\tt numpy}
documentation defines a conservative numeric tolerance for underflow
$\tau = 10^{-12}$, which is a conservative estimate of the numerical
stability boundary demonstrated in {\bf
  Figure~\ref{figure:badErrorsForAllPStar}} (those boundary points
occur very close to the true machine precision $\epsilon \approx
{10}^{-15}$).

Because Cooley-Tukey implementations of FFT-based convolution
(\emph{e.g.}, the {\tt numpy} implementation) are widely applied to
large problems with extremely small error, we will make a
simplification and assume that, when constraining the FFT result to
reach a value higher than machine epsilon (+ tolerance threshold), the
error from the FFT is negligible in comparison to the error introduced
by the $p^*$-norm approximation. This is firstly because the only
source of numerical error during FFT (assuming an FFT implementation
with numerically precise twiddle factors) on vectors in ${[0,1]}^k$
will be the result of underflow from repeated addition and subtraction
(neglecting the non-influencing multiplication with twiddle factors,
which each have magnitude $1$). The numerically imprecise routines are
thus limited to $(x + y) - x$; when $x >> y$ (\emph{i.e.},
$\frac{y}{x} < \epsilon \approx {10}^{-15}$, the machine precision),
then $(x + y) - x$ will return $0$ instead of $y$. To recover at least
one bit of the significand, the intermediate results of the FFT
must surpass machine precision $\epsilon$ (since the worst case
addition initially happens with the maximum $x=1.0$).

The maximum sum of any values from a list of $k$ such elements can
never exceed $k$; for this reason, a conservative estimate of the
numerical tolerance of an FFT (with regard to underflow) will be the
smallest value of $y$ for which $\frac{y}{k} > \epsilon$; thus, $y >
\epsilon k$. This yields a conservative estimate of the minimum value
in one index at the result of an FFT convolution: when the result at
some index $m$ is $> \epsilon k$, then the result should be
numerically stable. For this reason, we use a numerical tolerance
$\tau = {10}^{-12}$, thereby ensuring that the vast majority of
numerical error for the numerical max-convolution algorithm is due to
the $p^*$-norm approximation (\emph{i.e.}, employing $\| u^{(m)}
\|_{p^*}$ instead of $\| u^{(m)} \|_\infty$) and not due to the
long-used and numerically performant FFT result. Furthermore, in
practice the mean squared error due to FFT will be much smaller than
the conservative worst-case outlined here, because it is difficult for
the largest intermediate summed value (in this case $x$) to be
consistently large when many such very small values (in this case $y$)
are encountered in the same list. Although $\tau$ could be chosen
specifically for a problem of size $k$, note that this simple
derivation is very conservative and thus it would be better to use a
tighter bound for choosing $\tau$. Regardless, for an FFT
implementation that isn't as performant (\emph{e.g.}, because it uses
{\tt float} types instead of {\tt double}), increasing $\tau$ slightly
would suffice.

Therefore, from this point forward we consider that the dominant cause
of error to come from the max-convolution approximation. Using larger
$p^*$ values will provide a closer approximation; however, using a
larger value of $p^*$ may also drive values to zero (because the
inputs $L$ and $R$ will be normalized within {\bf
  Algorithm~\ref{algorithm:numericalMaxConvolveGivenPStar}} so that
the maximum of each is 1 when convolved via FFT), limiting the
applicability of large $p^*$ to indices $m$ for which $vM[m] \geq
\tau$.

Through this lens, the choice of $p^*$ can be characterized by two
opposing sources of error: higher $p^*$ values better approximate $\|
u^{(m)} \|_{p^*}$ but will be numerically unstable for many indices;
lower $p^*$ values provide worse approximations of $\| u^{(m)}
\|_{p^*}$ but will be numerically unstable for only few indices. These
opposing sources of error pose a natural method for improving the
accuracy of this max-convolution approximation. By considering a small
collection of $p^*$ values, we can compute the full numerical estimate
(at all indices) with each $p^*$ using {\bf
  Algorithm~\ref{algorithm:numericalMaxConvolveGivenPStar}}; computing
the full result at a given $p^*$ is $\in O(k \log_2(k))$, so doing so
on some small number $c$ of $p^*$ values considered, then the overall
runtime will be $\in O(c k \log_2(k) )$. Then, a final estimate is
computed at each index by using the largest $p^*$ that is stable (with
respect to underflow) at that index. Choosing the largest $p^*$ (of
those that are stable with respect to underflow) corresponds to
minimizing the bleed-in error, because the larger $p^*$ becomes, the
more muted the non-maximal terms in the norm become (and thus the
closer the $p^*$-norm becomes to the true maximum).

Here we introduce this piecewise method and compare it to the simpler
low-value $p^*=8$ and high-value $p^*=64$ methods and analyze the
worst-case error of the piecewise method.

\begin{algorithm}
  \caption{ {\bf Piecewise numerical max-convolution }, a numerical
    method to estimate the max-convolution of nonnegative vectors
    (revised to reduce bleed-in error). This procedure uses a $p^*$
    close to the largest possible stable value at each result
    index. The return value is a numerical estimate of the
    max-convolution $L *_{\max} R$. The runtime is in $O(k \log_2(k)
    \log_2(p^*_{\max}))$.}

  \label{algorithm:numericalMaxConvolvePiecewise}
  \begin{small}
    \begin{algorithmic}[1]
      \Procedure{numericalMaxConvolvePiecewise}{$L$, $R$, $p^*_{\max}$}

      \State $\ell_{\max} \gets \argmax_\ell L[\ell]$
      \State $r_{\max} \gets \argmax_r R[r]$
      \State $L' \gets \frac{L}{L[\ell_{\max}]}$
      \State $R' \gets \frac{R}{R[r_{\max}]}$ \Comment{Scale to a proportional problem on $L',R'$}

      \State $allPStar \gets [ 2^0, 2^1,\dots, 2^{\ceil[\big]{\log_2(p^*_{\max})}} ]$
      
      \For{$i \in \{ 0, 1, \ldots len(allPStar)\}$}
      \State $resForAllPStar[i] \gets$ \texttt{fftNonnegMaxConvolveGivenPStar}($L'$, $R'$, $allPStar[i]$)
      \EndFor
      
      \For{$m \in \{ 0, 1, \ldots len(L)+len(R)-1\}$}
      \State $maxStablePStarIndex[m] \gets \max \{ i:~ {\left( resForAllPStar[i][m] \right)}^{\text{allPStar[$i$]}} \geq \tau) \}$
      \EndFor

      \For{$m \in \{ 0, 1, \ldots len(L)+len(R)-1\}$}
      \State $i \gets maxStablePStarIndex[m]$
      \State $result[m] \gets resForAllPStar[i][m]$
      \EndFor

      \State \Return $L[\ell_{\max}] \times R[r_{\max}] \times result$
      \Comment{Undo previous scaling}

      \EndProcedure
    \end{algorithmic}
  \end{small}
\end{algorithm}

\section{Results}

This section derives theoretical error bounds as well as a practical 
comparison on an example for the standard piecewise method. Furthermore
the development of an improvement with affine scaling is shown.
Eventually, an evaluation of the latter is performed on a larger problem. 
Therefore we applied our technique to compute the Viterbi path for a 
hidden Markov model (HMM) to assess runtime and the level of error propagation. 

\subsection{Error and Runtime Analysis of the Piecewise Method}

We first analyze the error for a particular underflow-stable $p^*$ and
then use that to generalize to the piecewise method, which seeks to
use the highest underflow-stable $p^*$.

\subsubsection{Error Analysis for a Fixed Underflow-Stable $p^*$:}
We first scale $L$ and $R$ into $L'$ and $R'$ respectively, where the
maximum elements of both $L'$ and $R'$ are $1$; the absolute error can
be found by unscaling the absolute error of the scaled problem:
\begin{multline*}
  | exact(L,R)[m] - numeric(L',R')[m] |\\
  = \max_\ell L[\ell] ~ \max_r R[r] \; | exact(L',R')[m] - numeric(L',R')[m] |.
\end{multline*}
We first derive an error bound for the scaled problem on $L', R'$ (any
mention of a vector $u^{(m)}$ refers to the scaled problem), and then
reverse the scaling to demonstrate the error bound on the original
problem on $L, R$.

For any particular ``underflow-stable'' $p^*$ (\emph{i.e.}, any value
of $p^*$ for which ${\left(\| u^{(m)} \|_{p^*}\right)}^{p^*} \geq
\tau$), the absolute error for the numerical method for fast
max-convolution can be bound fairly easily by factoring out the
maximum element of $u^{(m)}$ (this maximum element is equivalent to
the Chebyshev norm) from the $p^*$-norm:

\[| exact(L',R')[m] - numeric(L',R')[m] |\]
\begin{eqnarray*}
  &=& | \| u^{(m)} \|_{p^*} - \| u^{(m)} \|_\infty | \\
  &=& \| u^{(m)} \|_{p^*} - \| u^{(m)} \|_\infty\\
  &=& \| u^{(m)} \|_\infty \left( \frac{\| u^{(m)} \|_{p^*}}{\| u^{(m)} \|_\infty} - 1 \right)\\
  &=& \| u^{(m)} \|_\infty \left( \| \frac{u^{(m)}}{\| u^{(m)} \|_\infty} \|_{p^*} - 1 \right)\\
  &=& \| u^{(m)} \|_\infty \left( \| v^{(m)} \|_{p^*} - 1 \right)
\end{eqnarray*}

where $v^{(m)}$ is a nonnegative vector of the same length as
$u^{(m)}$ (this length is denoted $k_m$) where $v^{(m)}$ contains one
element equal to $1$ (because the maximum element of $u^{(m)}$ must,
by definition, be contained within $u^{(m)}$) and where no element of
$v^{(m)}$ is greater than 1 (also provided by the definition of the
maximum).
\begin{eqnarray*}
  \| v^{(m)} \|_{p^*} &\leq& \| (1, 1, \ldots 1) \|_{p^*}\\
  &=& {\left( \sum_i^{k_m} 1^{p^*} \right)}^{\frac{1}{p^*}}\\
  &=& {k_m}^{\frac{1}{p^*}}
\end{eqnarray*}

Thus, since $\| v^{(m)} \|_{p^*} \geq 1$, the error is bound:
\begin{eqnarray*}
  &&| exact(L',R')[m] - numeric(L',R')[m]\\
  &=& \| u^{(m)} \|_\infty \left(\| v^{(m)} \|_{p^*} - 1 \right)\\
  &\leq& \| v^{(m)} \|_{p^*} - 1\\
  &\leq& k_m^\frac{1}{p^*} - 1,\\
\end{eqnarray*}
because $\forall m,~ \| u^{(m)} \|_\infty \leq 1$ for a scaled problem
on $L', R'$.

\subsubsection{Error Analysis of Piecewise Method}
However, the bounds derived above are only applicable for $p^*$ where
$\| u^{(m)} \|_{p^*}^{p^*} \geq \tau$. The piecewise
method is slightly more complicated, and can be partitioned into two
cases: In the first case, the top contour is used (\emph{i.e.}, when
$p^*_{\max}$ is underflow-stable).
Conversely, in the second case, a
middle contour is used (\emph{i.e.}, when $p^*_{\max}$ is not
underflow-stable). In this context, in general a contour
comprises of a set of indices $m$ with the same maximum stable $p^*$.

In the first case, when we use the top contour $p^* = p^*_{\max}$, we
know that $p^*_{\max}$ must be underflow-stable, and thus we can reuse
the bound given an underflow-stable $p^*$.

In the second case, because the $p^*$ used is $< p^*_{\max}$, it
follows that the next higher contour (using $2 p^*$) must not be
underflow-stable (because the highest underflow-stable $p^*$ is used
and because the $p^*$ are searched in log-space). The bound derived above that demonstrated 
\[ \| u^{(m)} \|_{p^*} \leq \| u^{(m)} \|_\infty k_m^\frac{1}{p^*} \]
can be combined with the property that $\| \cdot \|_{p^*} \geq \|
\cdot \|_\infty$ for any $p^* \geq 1$ to show that
\[ \| u^{(m)} \|_\infty \in \left[ \frac{\| u^{(m)} \|_{p^*}}{k_m^\frac{1}{p^*}} , \| u^{(m)} \|_{p^*} \right]. \]
Thus the absolute error can be bound again using the fact that we are
in a middle contour:
\begin{eqnarray*}
  &=& \| u^{(m)} \|_{p^*} - \| u^{(m)} \|_\infty\\
  &=& \| u^{(m)} \|_{p^*} \left( 1 - \frac{\| u^{(m)} \|_\infty}{\| u^{(m)} \|_{p^*}} \right)\\
  &\leq& \| u^{(m)} \|_{p^*} \left( 1 - k_m^\frac{-1}{p^*} \right)\\
  &<& \tau^\frac{1}{2 p^*} \left( 1 - k_m^\frac{-1}{p^*} \right).\\
\end{eqnarray*}

The absolute error from middle contours will be quite small when $p^*
= 1$ is the maximum underflow-stable value of $p^*$ at index $m$,
because $\tau^\frac{1}{2 p^*}$, the first factor in the error bound,
will become $\sqrt{\tau} \approx 10^{-6}$, and $1 - k_m^\frac{-1}{p^*}
< 1$ (qualitatively, this indicates that a small $p^*$ is only used
when the result is very close to zero, leaving little room for
absolute error). Likewise, when a very large $p^*$ is used, then $1 -
k_m^\frac{-1}{p^*}$ becomes very small, while $\tau^\frac{1}{2 p^*} <
1$ (qualitatively, this indicates that when a large $p^*$ is used, the
$\| \cdot \|_{p^*} \approx \| \cdot \|_\infty$, and thus there is
little absolute error). Thus for the extreme values of $p^*$, middle
contours will produce fairly small absolute errors. The unique mode
$p^*_{mode}$ can be found by finding the value that solves
\[ \frac{\partial}{\partial p^*_{mode}} ~ \left( \tau^\frac{1}{2 p^*_{mode}} \left( 1 - k_m^\frac{-1}{p^*_{mode}} \right) \right) = 0,\]
which yields
\[p^*_{mode} = \frac{\log_2(k_m)}{\log_2( -\frac{2 \log_2(k) - \log_2(\tau)}{\log_2(\tau)} )}.\]

An appropriate choice of $p^*_{\max}$ should be $> p^*_{mode}$ so that
the error for any contour (both middle contours and the top contour)
is smaller than the error achieved at $p^*_{mode}$, allowing us to
use a single bound for both. Choosing $p^*_{\max} = p^*_{mode}$ would
guarantee that all contours are no worse than the middle-contour error
at $p^*_{mode}$; however, using $p^*_{\max} = p^*_{mode}$ is still
quite liberal, because it would mean that for indices in the highest
contour (there must be a nonempty set of such indices, because the
scaling on $L'$ and $R'$ guarantees that the maximum index will have
an exact value of $1$, meaning that the approximation endures no
underflow and is underflow-stable for every $p^*$), a better error
\emph{could} be achieved by increasing $p^*_{\max}$. For this reason,
we choose $p^*_{\max}$ so that the top-contour error produced at
$p^*_{\max}$ is not substantially larger than all errors produced for
$p^*$ before the mode (\emph{i.e.}, for $p^* < p^*_{mode}$).

Choosing any value of $p^*_{\max} > p^*_{mode}$ guarantees the
worst-case absolute error bound derived here; however, increasing
$p^*_{\max}$ further over $p^*_{mode}$ may possibly improve the mean
squared error in practice (because it is possible that many indices in
the result would be numerically stable with $p^*$ values substantially
larger than $p^*_{mode}$). However, increasing $p^*_{\max} >>
p^*_{mode}$ will produce diminishing returns and generally benefit
only a very small number of indices in the result, which have exact
values very close to $1$. In order to balance these two aims
(increasing $p^*_{\max}$ enough over $p^*_{mode}$ but not excessively
so), we make a qualitative assumption that a non-trivial number of
indices require us to use a $p^*$ below $p^*_{mode}$; therefore,
increasing $p^*_{\max}$ to produce an error significantly smaller than
the lowest worst-case error for contours below the mode (\emph{i.e.}
$p^* < p^*_{mode}$) will increase the runtime without significantly
decreasing the mean squared error (which will become dominated by the
errors from indices that use $p^* < p^*_{mode}$). The lowest
worst-case error contour below the mode is $p^*=1$ (because the
absolute error function is unimodal, and thus must be increasing until
$p^*_{mode}$ and decreasing afterward); therefore, we heuristically
specify that $p^*_{\max}$ should produce a worst-case error on a
similar order of magnitude to the worst-case error produced with
$p^*=1$. In practice, specifying the errors at $p^*_{\max}$ and
$p^*=1$ should be equal is very conservative (it produces very large
estimates of $p^*_{\max}$, which sometimes benefit only one or two
indices in the result); for this reason, we heuristically choose that
the worst-case error at $p^*_{\max}$ should be no worse than square
root of the worst case error at $p^*=1$ (this makes the choice of
$p^*_{\max}$ less conservative because the errors at $p^*=1$ are very
close to zero, and thus their square root is larger). The square root
was chosen because it produced, for the applications described in this
paper, the smallest value of $p^*_{\max}$ for which the mean squared
error was significantly lower than using $p^*_{\max} = p^*_{mode}$
(the lowest value of $p^*_{\max}$ guaranteed to produce the absolute
error bound). This heuristic does satisfy the worst-case bound
outlined here (because, again, $p^*_{\max} > p^*_{mode}$), but it
could be substantially improved if an expected distribution of
magnitudes in the result vector were known ahead of time: prior
knowledge regarding the number of points stable at each $p^*$
considered would enable a well-motivated choice of $p^*_{\max}$ that
truly optimizes the expected mean squared error.

From this heuristic choice of $p^*_{\max}$, solving
\[ \sqrt{\sqrt{\tau} \left(1 - \frac{1}{k}\right)} = k^\frac{1}{p^*_{\max}} - 1 \]
(with the square root of the worst-case at $p^*=1$ on the left and the worst-case
error at $p^*_{\max}$ on the right) yields
\begin{eqnarray*}
p^*_{\max} &=& \frac{\log_2(k)}{\log_2(1+\sqrt{ \sqrt{\tau} \left( 1 - \frac{1}{k} \right) })}\\
&\approx& \frac{\log_2(k)}{\log_2(1+\sqrt{ \sqrt{\tau} })}
\end{eqnarray*}
for any non-trivial problem (\emph{i.e.}, when $k >> 1$), and thus
\[p^*_{\max} \approx \log_{1 + \tau^\frac{1}{4}}(k),\]
indicating that the absolute error at the top contour will be roughly
equal to the fourth root of $\tau$.\newline

\subsubsection{Worst-case Absolute Error}
By setting $p^*_{\max}$ in this manner, we guarantee that the absolute
error at any index of any unscaled problem on $L,R$ is less than
\[ \max_\ell L[\ell] ~ \max_r R[r] ~ \tau^\frac{1}{2 p^*_{mode}} \left( 1 - k_m^\frac{-1}{p^*_{mode}} \right) \]
where $p^*_{mode}$ is defined above. The full formula for the
middle-contour error at this value of $p^*_{mode}$ does not simplify
and is therefore quite large; for this reason, it is not reported
here, but this gives a numeric bound of the worst case middle-contour
error that is bound in terms of the variable $k$ (and with no other
free variables).

\subsubsection{Runtime Analysis}

The piecewise method clearly performs $\log_2(p^*_{\max})$ FFTs (each
requiring $O(k \log_2(k))$ steps); therefore, since $p^*_{\max}$ is
chosen to be $\log_{1 + \tau^\frac{1}{4}}(k)$ (to achieve the desired
error bound), the total runtime is thus
\[ O(k \log_2(k) \log_2(\log_{1 + \tau^\frac{1}{4}}(k)).\]
For any practically sized problem, the $\log_2(\log_{1 +
  \tau^\frac{1}{4}}(k))$ factor is essentially a constant; even when
$k$ is chosen to be the number of particles in the observable universe
($\approx 2^{270}$; \citealp{Eddington1923}), the $\log_2(\log_{1 +
  \tau^\frac{1}{4}}(k))$ is $\approx 18$, meaning that for any problem
of practical size, the full piecewise method is no more expensive than
computing between $1$ and $18$ FFTs.

\subsection{Comparison of Low-Value $p^*=8$, High-value $p^*=64$, and Piecewise Method}

\begin{figure*}
\centering
\begin{subfigure}[t]{.48\textwidth}
  \centering
  \includegraphics[width=\linewidth]{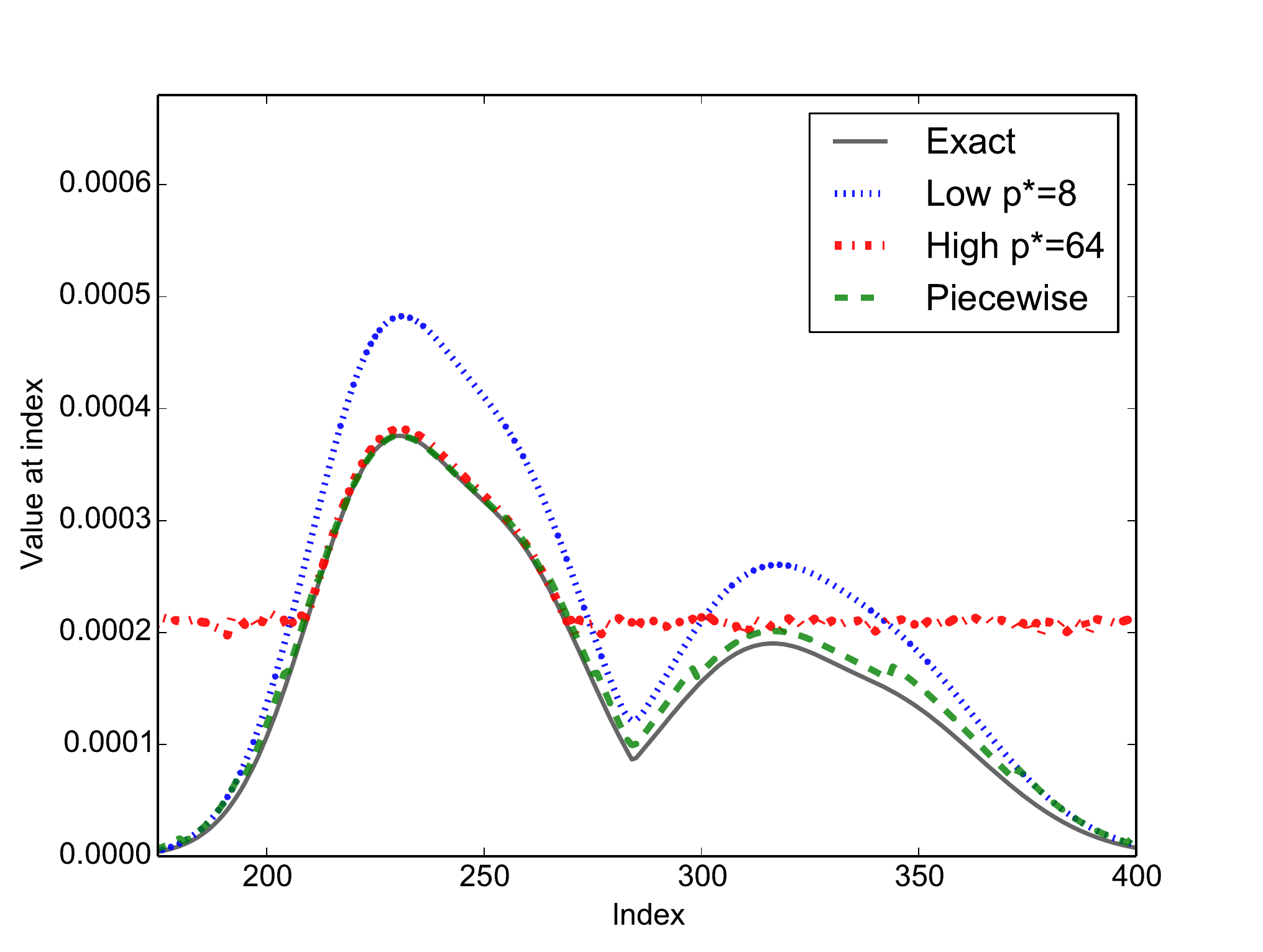}
  \caption{}
  \label{figure:doubleFigMethodsBimodalExa}
\end{subfigure}%
\hfill
\begin{subfigure}[t]{.48\textwidth}
  \centering
  \includegraphics[width=\linewidth]{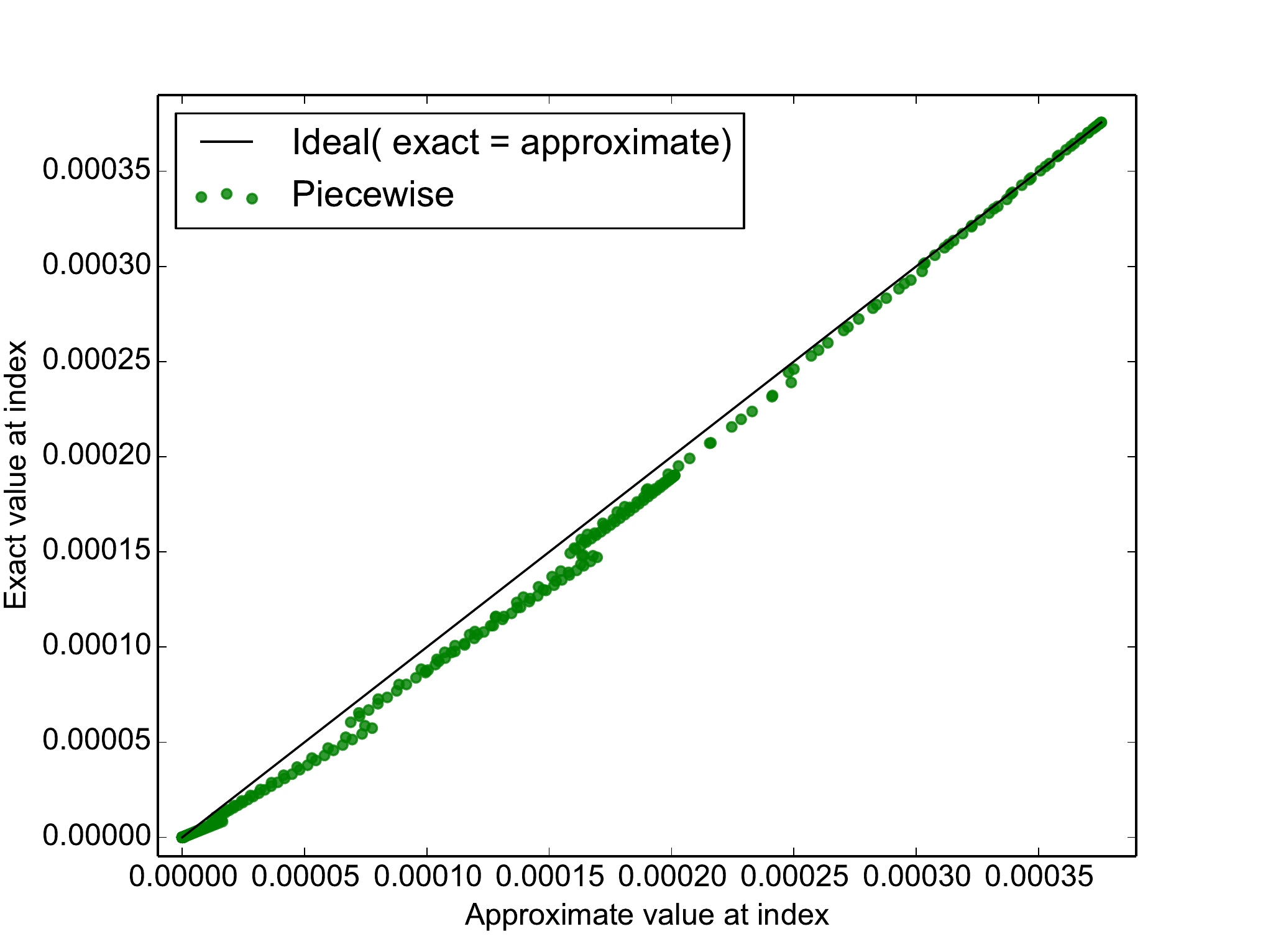}
  \caption{}
  \label{figure:doubleFigMethodsBimodalExb}
\end{subfigure}
\caption{{\bf The accuracy of numerical fast max-convolution methods.}
  {\bf (a)} Different approximations for a sample max-convolution
  problem. The low-$p^*$ method is underflow-stable, but overestimates
  the result. The high-$p^*$ method is accurate when underflow-stable,
  but experiences underflow at many indices. The piecewise method
  stitches together approximations from different $p^*$ to maintain
  underflow-stability. {\bf (b)} Exact vs. piecewise approximation at
  various indices of the same problem. A clear banding pattern is
  observed with one tight, elliptical cluster for each contour. The
  slope of the clusters deviates more for the contours using lower
  $p^*$ values.}
\label{figure:doubleFigMethodsBimodalEx}
\end{figure*}

We first use an example max-convolution problem to compare the results
from the low-value $p^*=8$, the high-value $p^*=64$ and piecewise
methods. At every index, these various approximation results are
compared to the exact values, as computed by the naive quadratic
method ({\bf Figure~\ref{figure:doubleFigMethodsBimodalExa}}).

\subsection{Improved Affine Piecewise Method}

\begin{figure}
\centering
\includegraphics[width=5in]{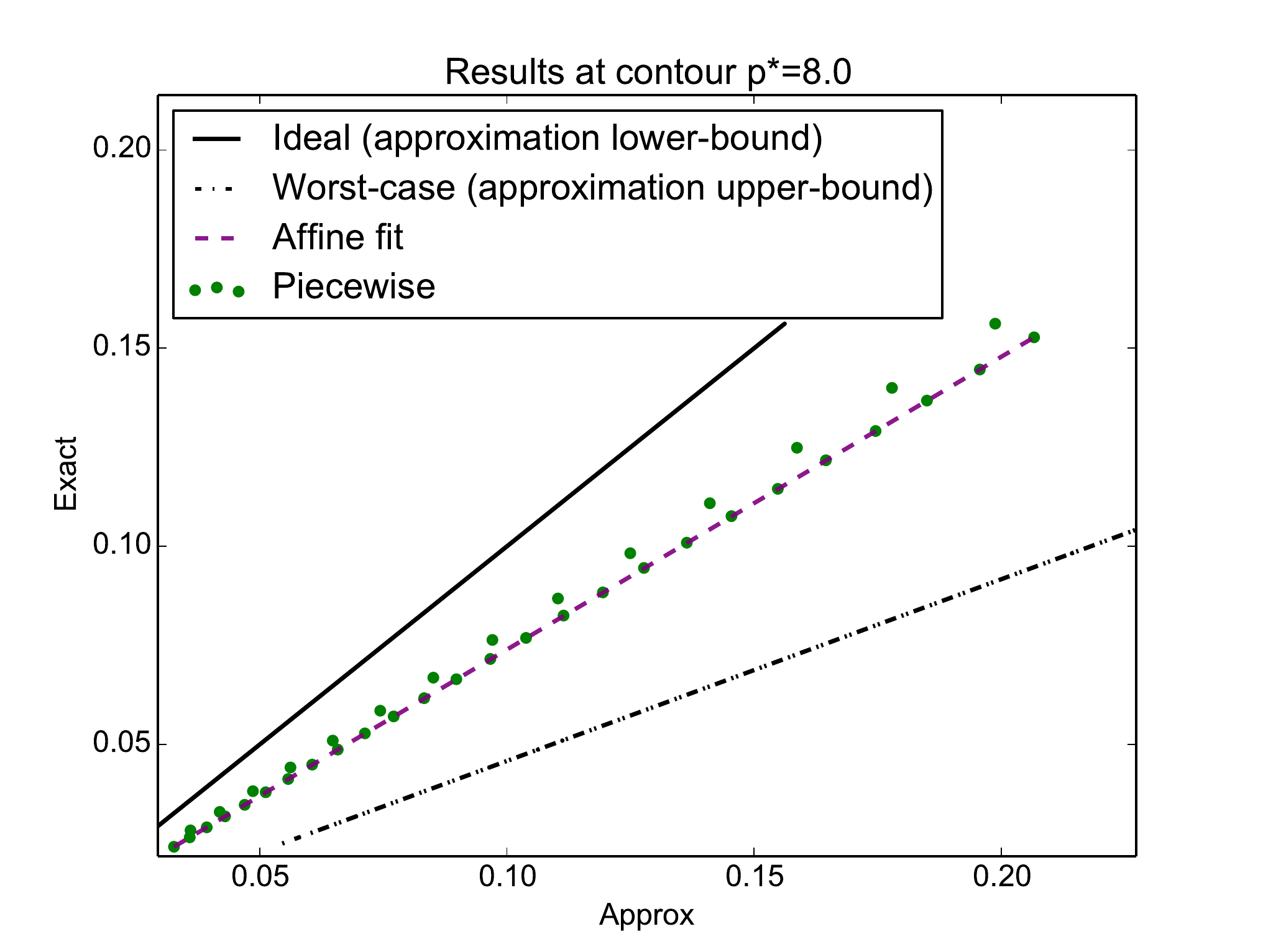}
\caption{{\bf A single contour from the piecewise approximation.} The
  cluster of points (one point for each index in the previous figure)
  is bounded by the exact value (ideal approximation) and the
  approximation upper-bound for $p^*=8$ (worst-case
  approximation). The points are well described by an affine function
  fit using the left-most and right-most points.}
\label{figure:approxExactLinearZoom}
\end{figure}

{\bf Figure~\ref{figure:doubleFigMethodsBimodalExb}} depicts a scatter
plot of the exact result vs. the piecewise approximation at every
index (using the same problem from {\bf
  Figure~\ref{figure:doubleFigMethodsBimodalExa}}). It shows a clear
banding pattern: the exact and approximate results are clearly
correlated, but each contour (\emph{i.e.}, each collection of indices
that use a specific $p^*$) has a different average slope between the
exact and approximate values, with higher $p^*$ contours showing a
generally larger slope and smaller $p^*$ contours showing greater
spread and lower slopes. This intuitively makes sense, because the
bounds on $\| u^{(m)} \|_\infty \in [ \| u^{(m)} \|_{p^*}
  k_m^\frac{-1}{p^*}, \| u^{(m)} \|_{p^*} ]$ derived above constrain
the scatter plot points inside a quadrilateral envelope ({\bf
  Figure~\ref{figure:approxExactLinearZoom}}).

The correlations within each contour can be exploited to correct
biases that emerge for smaller $p^*$ values. In order to do this, $\|
u^{(m)} \|_\infty$ must be computed for at least two points $m_1$ and
$m_2$ within the contour, so that a mapping $\| u^{(m)} \|_{p^*}
\approx f(\| u^{(m)} \|_{p^*}) = a \| u^{(m)} \|_{p^*} + b$ can be
constructed. Fortunately, a single $\| u^{(m)} \|_\infty$ can be
computed exactly in $O(k)$ (by actually computing a single $u^{(m)}$
and computing its max, which is equivalent to computing a single index
result via the naive quadratic method). As long as the exact value $\|
u^{(m)} \|_\infty$ is computed for only a small number of indices,
the order of the runtime will not change (each contour already
costs $O(k \log_2(k))$, so adding a small number of $O(k)$ steps for
each contour will not change the asymptotic runtime). 

If the two indices chosen are 
\[ m_{\min} = \argmin_{m \in contour(p^*)} \| u^{(m)} \|_{p^*} \]
\[ m_{\max} = \argmax_{m \in contour(p^*)} \| u^{(m)} \|_{p^*},\]
then we are guaranteed that the affine function $f$ can be written as
a convex combination of the exact values at those extreme points
(using barycentric coordinates):

\[ f(\| u^{(m)} \|_{p^*}) = \lambda_m \| u^{(m_{\max})} \|_\infty + \left( 1- \lambda_m \right) \| u^{(m_{\min})} \|_\infty \]
\[ \lambda_m = \frac{\| u^{(m)} \|_{p^*} - \| u^{(m_{\min})} \|_{p^*}}{\| u^{(m_{\max})} \|_{p^*} - \| u^{(m_{\min})} \|_{p^*}} \in [0,1]\]

Thus, by computing $\| u^{(m_{\min})} \|_\infty$ and $\|
u^{(m_{\max})} \|_\infty$ (each in $O(k)$ steps), we can compute an
affine function $f$ to correct contour-specific trends ({\bf
  Algorithm~\ref{algorithm:numericalMaxConvolvePiecewiseImproved}}).

\begin{figure*}
\centering
\begin{subfigure}[t]{.48\textwidth}
  \centering
  \includegraphics[width=\linewidth]{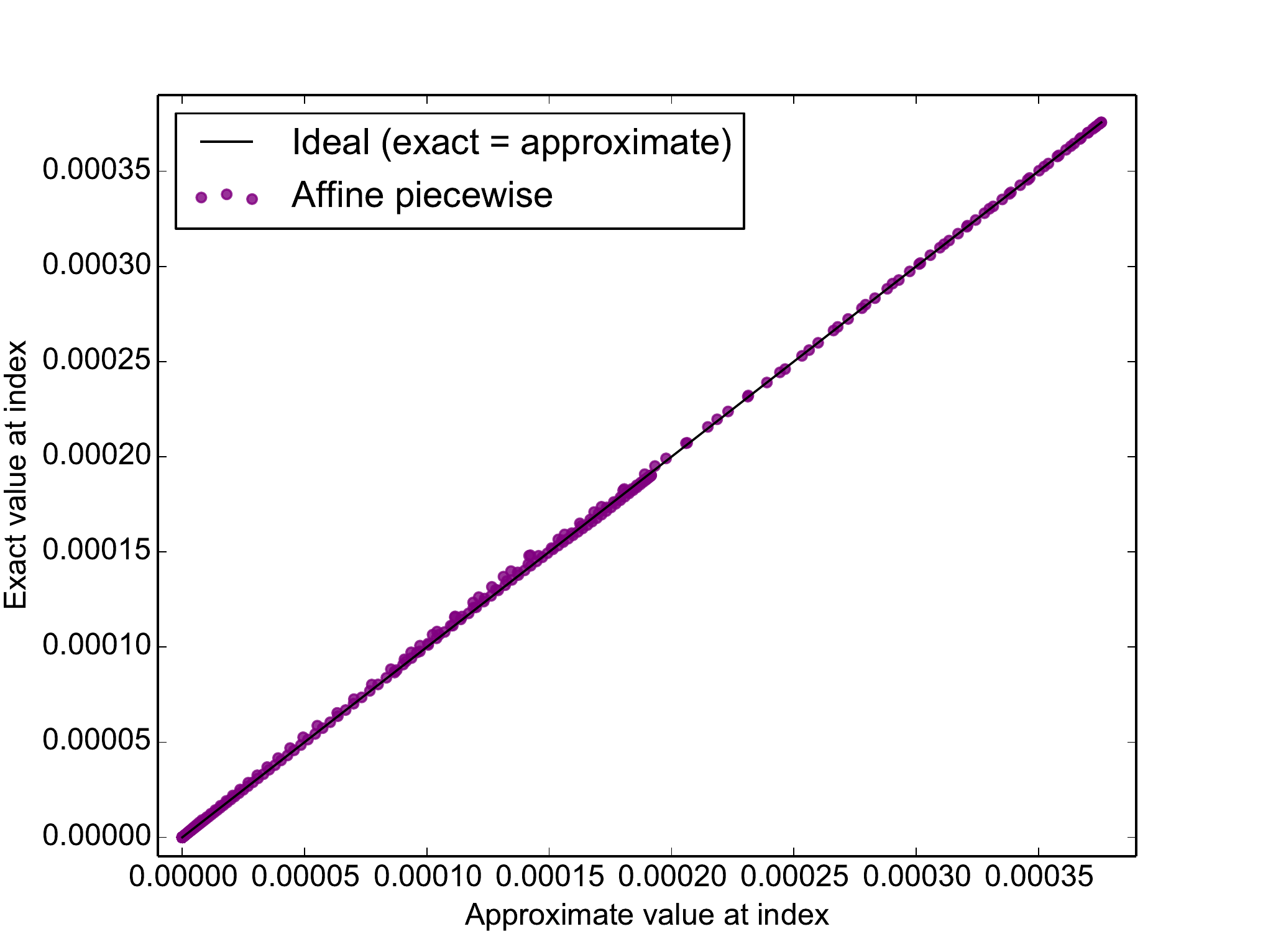}
  \caption{}
  \label{figure:approxExactLinearA}
\end{subfigure}%
\hfill
\begin{subfigure}[t]{.48\textwidth}
  \centering
  \includegraphics[width=\linewidth]{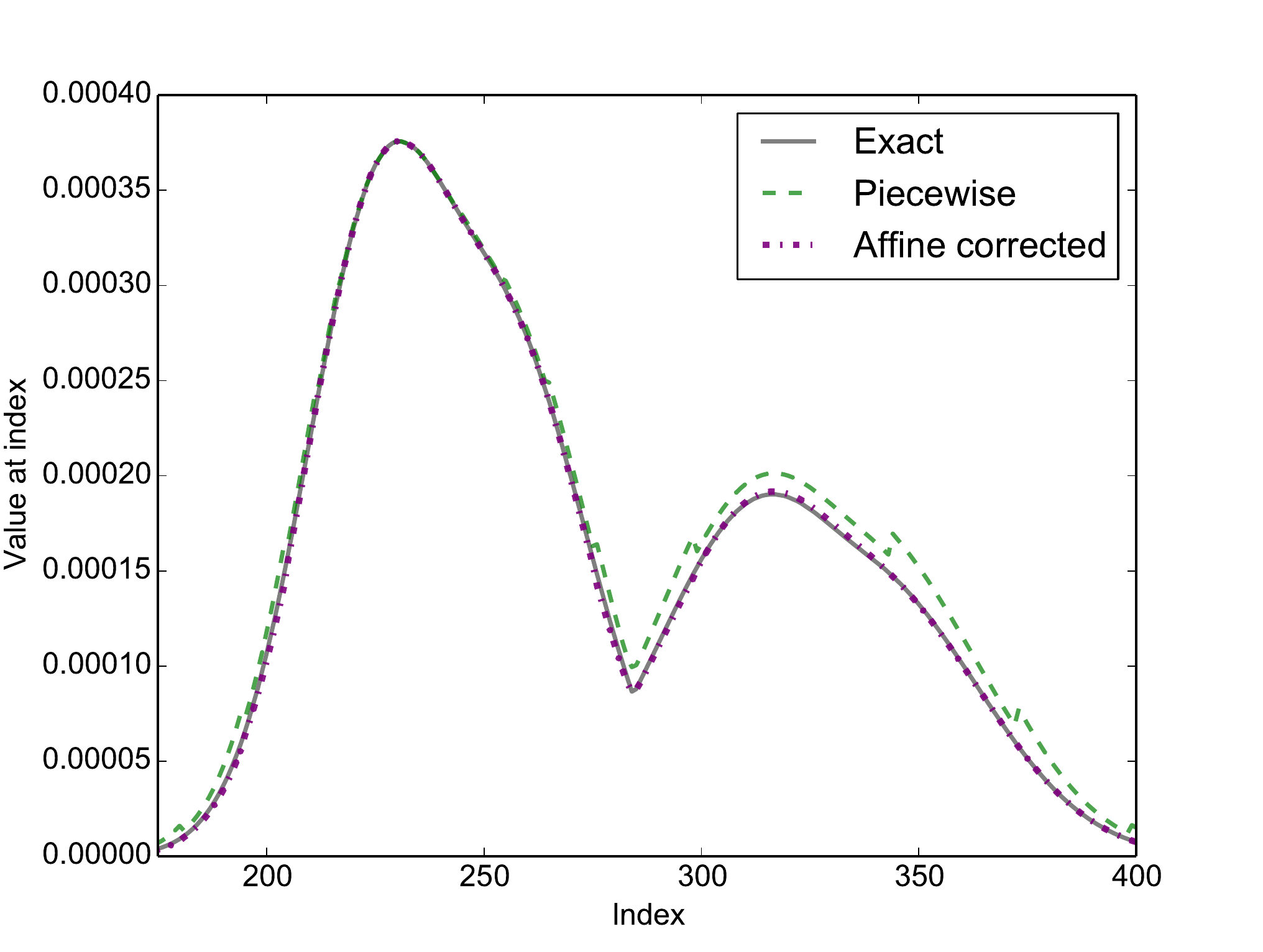}
  \caption{}
  \label{figure:approxExactLinearB}
\end{subfigure}
\caption{{\bf Piecewise method with affine contour fitting.} The
  approximate values at each index of the max-convolution problem are
  almost identical to the exact result at the same index.}
\label{figure:approxExactLinear}
\end{figure*}

\begin{algorithm}
  \caption{ {\bf Improved affine piecewise numerical
      max-convolution}, a numerical method to estimate the
    max-convolution nonnegative vectors (further revised to reduce
    numerical error). This procedure uses a $p^*$ close to the largest
    possible stable value at each result index. The return value is a
    numerical estimate of the max-convolution $L *_{\max} R$. The
    runtime is in $O(k \log_2(k) \log_2(p^*_{\max}))$.}

  \label{algorithm:numericalMaxConvolvePiecewiseImproved}
  \begin{small}
    \begin{algorithmic}[1]
      \Procedure{numericalMaxConvolvePiecewiseAffine}{$L$, $R$, $p^*_{\max}$}

      \State $\ell_{\max} \gets \argmax_\ell L[\ell]$
      \State $r_{\max} \gets \argmax_r R[r]$
      \State $L' \gets \frac{L}{L[\ell_{\max}]}$
      \State $R' \gets \frac{R}{R[r_{\max}]}$ \Comment{Scale to a proportional problem on $L',R'$}

      \State $allPStar \gets [ 2^0, 2^1,\dots, 2^{\ceil[\big]{\log_2(p^*_{\max})}} ]$
      
      \For{$i \in \{ 0, 1, \ldots len(allPStar)\}$}
      \State $resForAllPStar[i] \gets$ \texttt{fftNonnegMaxConvolveGivenPStar}($L'$, $R'$, $allPStar[i]$)
      \EndFor
      
      \For{$m \in \{ 0, 1, \ldots len(L)+len(R)-1\}$}
      \State $maxStablePStarIndex[m] \gets \max \{ i:~ {\left( resForAllPStar[i][m] \right)}^{\text{allPStar[$i$]}} \geq \tau) \}$
      \EndFor
      
      \State $result \gets $\texttt{affineCorrect}$(resForAllPStar, maxStablePStarIndex)$

      \State \Return $L[\ell_{\max}] \times R[r_{\max}] \times result$
      \Comment{Undo previous scaling}

      \EndProcedure
    \end{algorithmic}
  \end{small}
\end{algorithm}

\begin{algorithm}
  \caption{ {\bf Subroutine for correcting errors in a contour}, 
  with an affine transformation based on exact boundary points.
  It needs the results of the evaluation of the different
  $p$-norms as well as the (index of the) maximum stable values of $p^*$
  at every index.}

  \label{algorithm:affineCorrectSubroutine}
  \begin{small}
    \begin{algorithmic}[1]
      \Procedure{affineCorrect}{$resForAllPStar$, $maxStablePStarIndex$}
      \State $\forall i, slope[i] \gets 1$
      \State $\forall i, bias[i] \gets 0$

      \State $usedPStar \gets set(maxStablePStarIndex)$
      \For{$i \in usedPStar$}
      \State $contour \gets \{ m : maxStablePStarIndex[m] = i \}$
      \State $mMin \gets argmin_{m \in contour} resForAllPStar[i][m]$
      \State $mMax \gets argmax_{m \in contour} resForAllPStar[i][m]$

      \State $xMin \gets resForAllPStar[i][mMin]$
      \State $xMax \gets resForAllPStar[i][mMax]$
      \State $yMin \gets \texttt{maxConvolutionAtIndex}(mMin)$
      \State $yMax \gets \texttt{maxConvolutionAtIndex}(mMax)$
      \If{$xMax > xMin$}
      \State $slope[i] \gets \frac{yMax-yMin}{xMax-xMin}$
      \State $bias[i] \gets yMin - slope[i] \times xMin$
      \Else
      \State $slope[i] \gets \frac{yMax}{xMax}$
      \EndIf
      \EndFor

      \For{$m \in \{ 0, 1, \ldots len(L)+len(R)-1\}$}
      \State $i \gets maxStablePStarIndex[m]$
      \State $result[m] \gets resForAllPStar[i][m] \times slope[i] + bias[i]$
      \EndFor
      
      \State \Return $result$
      \EndProcedure
    \end{algorithmic}
  \end{small}
\end{algorithm}
\subsubsection{Error Analysis of Improved Affine Piecewise Method}

By exploiting the convex combination used to define $f$, the absolute
error of the affine piecewise method can also be bound. Qualitatively,
this is because, by fitting on the extrema in the contour, we are now
interpolating. If the two points used to determine the parameters of
the affine function were not chosen in this manner to fit the affine
function, then it would be possible to choose two points with very
close x-values (\emph{i.e.}, similar approximate values) and disparate
y-values (\emph{i.e.}, different exact values), and extrapolating to other
points could propagate a large slope over a large distance; using the
extreme points forces the affine function to be a convex combination
of the extrema, thereby avoiding this problem.

\begin{multline*}
f(\| u^{(m)} \|_{p^*}) = \lambda_m \| u^{(m_{\max})} \|_\infty + \left( 1- \lambda_m \right) \| u^{(m_{\min})} \|_\infty\\
\shoveleft{ \in  \left[ \lambda_m \frac{\| u^{(m_{\max})} \|_{p^*}}{k_{m_{\max}}^\frac{1}{p^*}} + \left( 1 - \lambda_m \right) \frac{\| u^{(m_{\min})} \|_{p^*}}{k_{m_{\min}}^\frac{1}{p^*}} ,\right.}\\ \left. \lambda_m \| u^{(m_{\max})} \|_{p^*} + \left( 1 - \lambda_m \right) \| u^{(m_{\min})} \|_{p^*} \right]\\
\shoveleft{\subseteq  \left[ \lambda_m \frac{\| u^{(m_{\max})} \|_{p^*}}{k^\frac{1}{p^*}} + \left( 1 - \lambda_m \right) \frac{\| u^{(m_{\min})} \|_{p^*}}{k^\frac{1}{p^*}} ,\right.}\\ \left. \lambda_m \| u^{(m_{\max})} \|_{p^*} + \left( 1 - \lambda_m \right) \| u^{(m_{\min})} \|_{p^*} \right]\\
\shoveleft{ = \left[ k^\frac{-1}{p^*} \left( \lambda_m \| u^{(m_{\max})} \|_{p^*} + \left( 1 - \lambda_m \right) \| u^{(m_{\min})} \|_{p^*} \right) ,\right.}\\ \left. \lambda_m \| u^{(m_{\max})} \|_{p^*} + \left( 1 - \lambda_m \right) \| u^{(m_{\min})} \|_{p^*} \right]\\
\shoveleft{= \left[ k^\frac{-1}{p^*} \| u^{(m)} \|_{p^*}, \| u^{(m)} \|_{p^*} \right]}
\end{multline*}

The worst-case absolute error of the scaled problem on $L', R'$ can be defined
\[ \max_m | ~ f(\| u^{(m)} \|_{p^*}) - \| u^{(m)} \|_\infty ~ |. \]
Because the function $f(\| u^{(m)} \|_{p^*}) - \| u^{(m)} \|_\infty$
is affine, it's derivative can never be zero, and thus Lagrangian
theory states that the maximum must occur at a boundary
point. Therefore, the worst-case absolute error is
\begin{eqnarray*}
  & \leq & \max \{ \| u^{(m)} \|_{p^*} - \| u^{(m)} \|_\infty , \| u^{(m)} \|_\infty - \| u^{(m)} \|_{p^*} k^\frac{-1}{p^*} \}\\
  & = & \| u^{(m)} \|_{p^*} - \| u^{(m)} \|_\infty,\\
\end{eqnarray*}
which is identical to the worst-case error bound before applying the
affine transformation $f$. Thus applying the affine transformation can
dramatically improve error, but will not make it worse than the
original worst-case.

\subsection{Demonstration on Hidden Markov Model With Toeplitz Transition Matrix}
One example that profits from fast max-convolution of non-negative
vectors is computing the Viterbi path using a hidden Markov model
(HMM) (\emph{i.e.}, the \emph{maximum a posteriori} states) with an
additive transition function satisfying $\Pr(X_{i+1} = a | X_{i} = b)
\propto \delta(a-b)$ for some arbitrary function $\delta$ ($\delta$
can be represented as a table, because we are considering all possible
discrete functions). This additivity constraint is equivalent to the
transition matrix being a ``Toeplitz matrix'': the transition matrix
$T_{a,b} = \Pr(X_{i+1} = a | X_{i} = b)$ is a Toeplitz matrix when all
cells diagonal from each other (to the upper left and lower right)
have identical values (\emph{i.e.}, $\forall a, \forall b, ~ T_{a,b} =
T_{a+1,b+1}$). Because of the Markov property of the chain, we only
need to max-marginalize out the latent variable at time $i$ to compute
the distribution for the next latent variable $X_{i+1}$ and all
observed values of the data variables $D_0 \dots D_{i+1}$. This
procedure, called the Viterbi algorithm, is continued inductively:

\begin{multline*}
\shoveleft{\max_{x_0, x_1, \ldots x_{i-1}} \Pr(D_0, D_1, \ldots D_{i-1}, X_0=x_0, X_1=x_1, \ldots, X_i=x_i) =} \\
\\
\shoveleft{\max_{x_{i-1}} \max_{x_0, x_1, \dots x_{i-2}}~ \Pr(D_0, D_1, \ldots D_{i-2}, X_0=x_0, X_1=x_1, \ldots, X_{i-1}=x_{i-1})}\\
  \shoveright{\Pr(D_{i-1} | X_{i-1}=x_{i-1}) \Pr(X_i=x_i | X_{i-1}=x_{i-1})}\\
\end{multline*}
and continuing by exploiting the self-similarity on a smaller problem
to proceed inductively, revealing a max-convolution (for this
specialized HMM with additive transitions):
\begin{multline*}
\shoveleft{= \max_{x_{i-1}} ~ fromLeft[i-1] \Pr(D_{i-1} | X_{i-1}=x_{i-1}) \delta[x_i - x_{i-1}] =}\\
\shoveright{\left(fromLeft[i-1]~likelihood[D_{i-1}] \right) ~*_{\max}~ \delta[x_i - x_{i-1}].}\\
\end{multline*}

After computing this left-to-right pass (which consisted of $n-1$
max-convolutions and vector multiplications), we can find the
\emph{maximum a posteriori} configuration of the latent variables
$X_0,\ldots X_{n-1} = x_0^*,\ldots x_{n-1}^*$ backtracking
right-to-left, which can be done by finding the variable value $x_i$
that maximizes $fromLeft[i][x_i] \times \delta[x_{i+1}^* - x_i]$ (thus
defining $x_i^*$ and enabling induction on the right-to-left
pass). The right-to-left pass thus requires $O(n k)$ steps ({\bf
  Algorithm~\ref{algorithm:viterbi}}). Note that the full max-marginal
distributions on each latent variable $X_i$ can be computed via a
small modification, which would perform a more complex right-to-left
pass that is nearly identical to the left-to-right pass, but which
performs subtraction instead of addition (\emph{i.e.}, by reversing
the vector representation of the PMF of the subtracted argument before
it is max-convolved;~\citealp{Serang2014b}).

\begin{algorithm}
  \caption{ {\bf Viterbi for models with additive transitions}, which
    accepts the length $k$ vector $prior$, a list of $n$ binned
    observations $data$, a $a \times k$ matrix of likelihoods (where
    $a$ is the number of bins used to discretize the data)
    $likelihoods$, and a length $2 k-1$ vector $\delta$ that describes
    the transition probabilities. The algorithm returns a Viterbi path
    of length $n$, where each element in the path is a valid state
    $\in \{ 0, 1, \ldots k-1 \}$.}

  \label{algorithm:viterbi}
  \begin{small}
    \begin{algorithmic}[1]
      \Procedure{ViterbiForAdditiveTransitions}{$prior, data, likelihood, \delta$}
	  \State $fromLeft[0] \gets prior$
      \For {$i=0$ to $n-2$}
      	\State $fromLeft[i] \gets fromLeft[i] \times likelihood[data[i]]$
      	\State $fromLeft[i+1] \gets fromLeft[i] *_{\max} \delta$
      \EndFor 
	  \State $fromLeft[n] \gets fromLeft[n] \times likelihood[data[n]]$
      \State
      \State $path[n-1] \gets \argmax_{j} fromLeft[n-1][j]$
      \For {$i=n-2$ to $0$}
        \State $maxProdPosterior \gets -1$
        \State $argmaxProdPosterior \gets -1$
        \For {$l=k$ to $1$}
      		\State $currProdPosterior \gets fromLeft[i] \times \delta[l-path[i+1]]$
      		\If {$currProdPosterior>maxProdPosterior$}
      			\State $maxProdPosterior \gets currProdPosterior$
      			\State $argmaxProdPosterior \gets l$
      		\EndIf
      	\EndFor
      	\State $path[i] \gets argmaxProdPosterior$
      \EndFor
      \State \Return $path$
      \EndProcedure
    \end{algorithmic}
  \end{small}
\end{algorithm}

We apply this HMM with additive transition probabilities to a data
analysis problem from economics. It is known for example, that the
current figures of unemployment in a country have (among others)
impact on prices of commodities like oil. If one could predict
unemployment figures before the usual weekly or monthly release by the
responsible government bureaus, this would lead to an information
advantage and an opportunity for short-term arbitrage. The close
relation of economic indicators like market prices and stock market
indices (especially of indices combining several stocks of different
industries) to unemployment statistics can be used to tackle this
problem.

In the following demonstration of our method, we create a simple HMM
with additive transitions and use it to infer the \emph{maximum a
  posteriori} unemployment statistics given past history (\emph{i.e.}
how often unemployment is low and high, as well as how often
unemployment goes down or up in a short amount of time) and current
stock market prices (the observed data). We discretized random
variables for the observed data (S\&P 500, adjusted closing prices ;
retrieved from YAHOO! historical stock prices:
\url{http://data.bls.gov/cgi-bin/surveymost?bls series CUUR0000SA0}),
and "latent" variables (unemployment insurance claims, seasonally
adjusted, were retrieved from the U.S. Department of Labor:
\url{https://www.oui.doleta.gov/unemploy/claims.asp}). Stock prices
were additionally inflation adjusted by (\emph{i.e.} divided by) the
consumer price index (CPI) (retrieved from the U.S. Bureau of Labor
Statistics: \url{https://finance.yahoo.com/q?s=^GSPC}). The
intersection of both "latent" and observed data was available weekly
from week 4 in 1967 to week 52 in 2014, resulting in 2500 data points
for each type of variable.

\begin{figure*}
\centering
\includegraphics[width=5in]{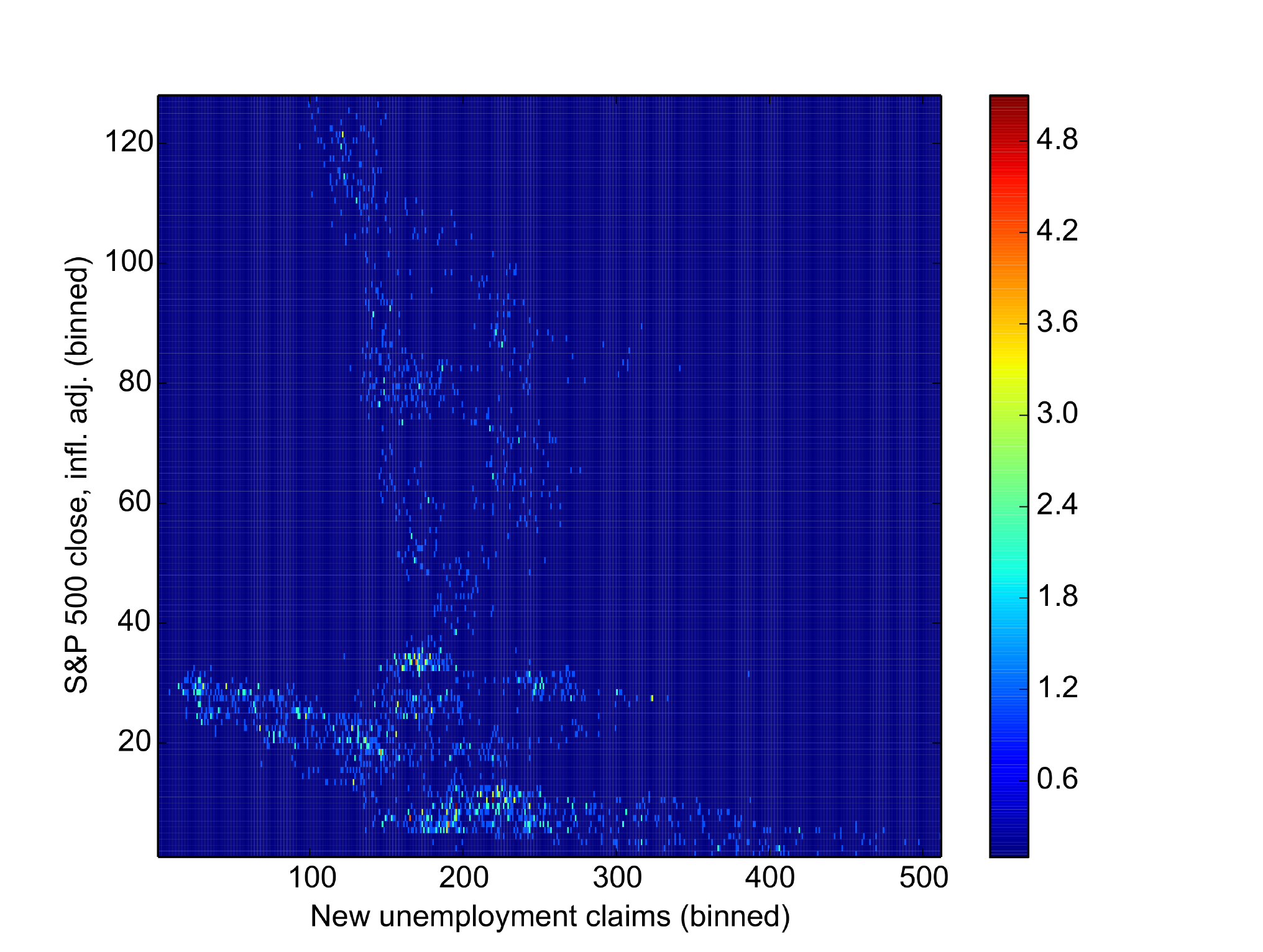}
\caption{{\bf Heatmap for trained likelihood matrix.} This heatmap
  depicts a joint empirical distribution between the S\&P 500 index
  and new unemployment claims, which share a tenuous inverse
  relationship. Given $D_i$, the discretized stock index value at time
  $i$, row $D_i$ contains the likelihood table $\Pr(D_i | X_i)$, which
  is denoted $likelihood[data[i]]$ in the code.}
\label{figure:likelihoodHeatmap}
\end{figure*}

To investigate the influence of overfitting, we partition the data in
two parts, before June 2005 and after June 2005, so that we are
effectively training on $\frac{2000 \times 100}{2500} = 80\%$ of the data points, and
then demonstrate the Viterbi path on the entirety of the data (both
the $80\%$ training data and the $20\%$ of the data withheld from
empirical parameter estimation). Unemployment insurance claims were
discretized into $512$ and stock prices were discretized into $128$
bins. Simple empirical models of the prior distribution for
unemployment, the likelihood of unemployment given stock prices, and
the transition probability of unemployment were built as follows: The
initial or prior distribution for unemployment claims at $i=0$ was
calculated by marginalizing the time series of training data for the
claims (\emph{i.e.} counting the number of times any particular
unemployment value was reached over all possible bins). Our transition
function (the conditional probability $\Pr(X_{i+1} | X_i)$) similarly
counts the number of times each possible change $X_{i+1}-X_i \in
\{-511, -510, \ldots 511\}$ occurred over all available time
points. Interestingly, the resulting transition distribution roughly
resembles a Gaussian (but is not an exact Gaussian). This underscores
a great quality of working with discrete distributions: while
continuous distributions may have closed-forms for max-convolution
(which can be computed quickly), discrete distributions have the
distinct advantage that they can accurately approximate any smooth
distribution. Lastly, the likelihoods of observing a stock price given
the unemployment at the same time were trained using an empirical
joint distribution (essentially a heatmap), which is displayed in {\bf
  Figure~\ref{figure:likelihoodHeatmap}}.

\begin{figure}
\centering
\includegraphics[width=5in]{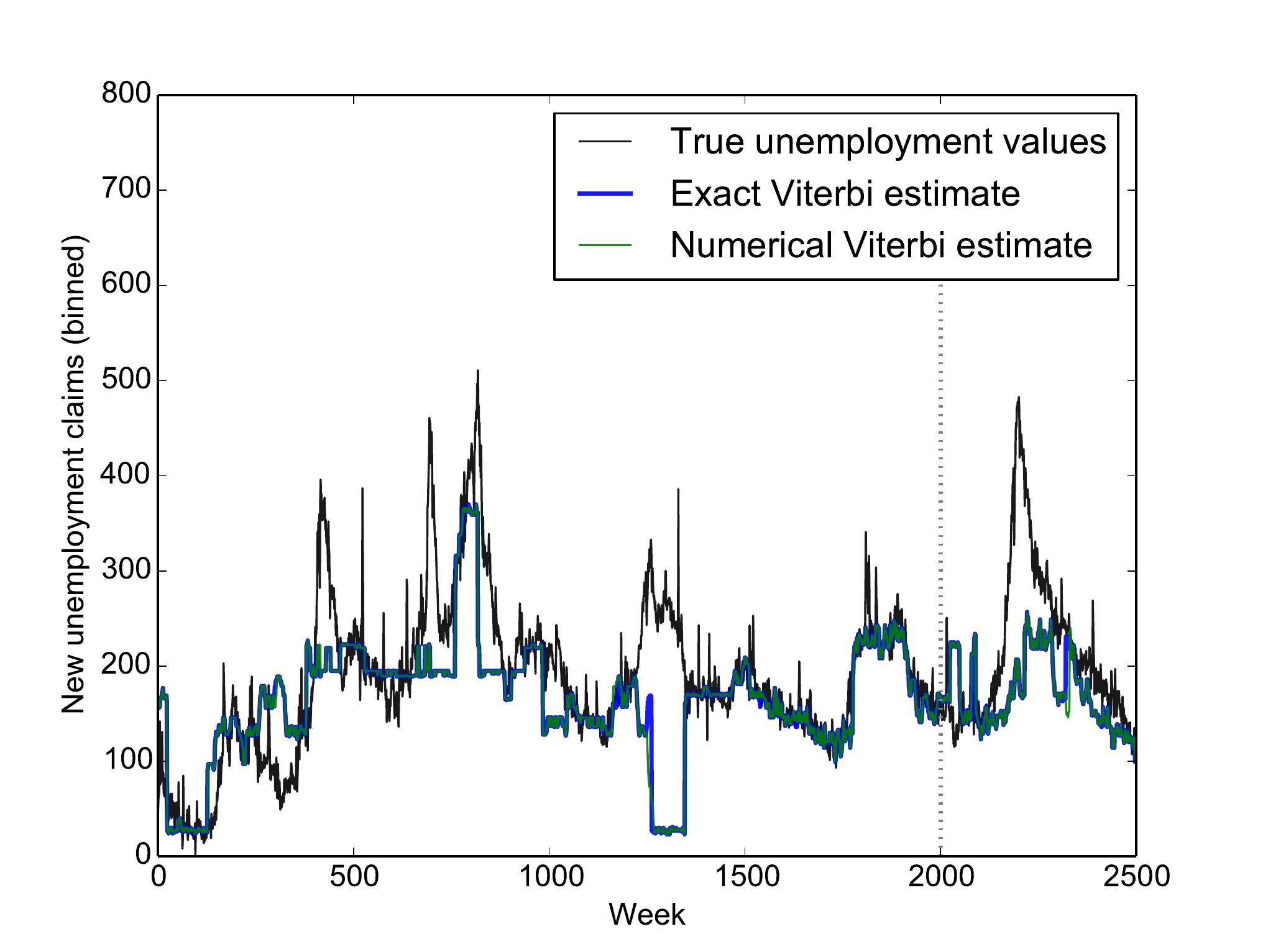}
\caption{{\bf Viterbi analysis of employment given stock index
    values.} The Viterbi path corresponding to the \emph{maximum a
    posteriori} prediction of the number of new unemployment insurance
  claims is produced for a model where the state transition
  probabilities are additive. The exact Viterbi estimate tracks well
  with the true unemployment values. Training parameters were taken
  from only the true unemployment data to the left of the vertical
  dotted line; however, the Viterbi paths to the right of the dotted
  line (where unemployment data were withheld from the likelihood,
  prior, and transition parameters) also track well with the true
  unemployment statistics. The Viterbi path computed with fast
  numerical max-convolution (via the affine piecewise approach) is
  nearly identical to the result computed with the slower exact
  approach.}
\label{figure:viterbi}
\end{figure}

We compute the Viterbi path two times: First we use naive, exact
max-convolution, which requires a total of $O(n k^2)$ steps. Second,
we use fast numerical max-convolution, which requires $O(n~k \log(k)
\log(\log(k))$ steps. Despite the simplicity of the model, the exact
Viterbi path (computed via exact max-convolution) is highly
informative for predicting the value of unemployment, even for the
$20\%$ of the data that were not used to estimate the empirical prior,
likelihood, and transition distributions. Also, the numerical
max-convolution method is nearly identical to the exact
max-convolution method at every index ({\bf
  Figure~\ref{figure:viterbi}}). Even with a fairly rough
discretization (\emph{i.e.}, $k=512$), the fast numerical method used
$141.4$ seconds compared to the $292.3$ seconds required by the naive
approach. This speedup will increase dramatically as $k$ is increased,
because the $\log(\log(k))$ term in the runtime of the numerical
max-convolution method is essentially bounded above $\log(\log(k))
\leq 18$.

\subsection{An Improved Approximation of the Chebyshev Norm}

Although the $p^*$-norm provides a good approximation of the Chebyshev
norm, it discards significant information; specifically the curve $\|
u^{(m)} \|_{p^*}$ for various ${p^*}$ could be used to identify and
correct the worst-case scenario where
$\frac{u^{(m)}}{\|u^{(m)}\|_\infty} = (1, 1, \ldots 1)$; using only
two points, the exact value of $\| u^{(m)} \|_\infty$ can be computed
for those worst-case $u^{(m)}$ vectors by computing the norms at two
different $p^*$ values and solving the following equations for
$\beta_1$:
\begin{eqnarray*}
\| u^{(m)} \|_{p^*_1}^{p^*_1} & \propto & \beta_1^{p^*_1}\\
\| u^{(m)} \|_{p^*_2}^{p^*_2} & \propto & \beta_1^{p^*_2},
\end{eqnarray*}
where the proportionality constant is $k_m = len(u^{(m)})$ and where
the computed value $\beta_1$ yields the exact Chebyshev norm $\|
u^{(m)} \|_\infty$.

\subsubsection{A Projection-Based Method for Estimating $\| u^{(m)} \|_\infty$}
More generally, when there are $e_m \leq k_m$ unique values ($\beta_i$) in
$u^{(m)}$, we can model the norms perfectly with
\[
\| u^{(m)} \|_{p^*}^{p^*} = \sum_i^{e_m} h_i \beta_i^{p^*}
\]
where $h_i$ is an integer that indicates the number of times $\beta_i$
occurs in $u^{(m)}$ (and where $\sum_i h_i = k_m =
len(u^{(m)})$). This multi-set view of the vector $u^{(m)}$ can be
used to project it down to a dimension $r$:
\[
\left[
\begin{array}{cccc}
\alpha_1^{p^*} & \alpha_2^{p^*} & & \alpha_r^{p^*} \\
\alpha_1^{2 {p^*}} & \alpha_2^{2 {p^*}} & \cdots & \alpha_r^{2 {p^*}} \\
\alpha_1^{3 {p^*}} & \alpha_2^{3 {p^*}} & & \alpha_r^{3 {p^*}} \\
\vdots & \vdots & & \vdots \\
\alpha_1^{\ell {p^*}} & \alpha_2^{\ell {p^*}} & & \alpha_r^{\ell {p^*}} \\
\end{array}
\right]
\cdot
\left[
\begin{array}{c}
n_1\\
n_2\\
n_3\\
\vdots\\
n_r\\
\end{array}
\right]
=
\left[
\begin{array}{c}
\| u^{(m)} \|_{p^*}^{p^*} \\
\| u^{(m)} \|_{2 {p^*}}^{2 {p^*}}\\
\| u^{(m)} \|_{2 {p^*}}^{2 {p^*}}\\
\vdots\\
\| u^{(m)} \|_{\ell {p^*}}^{\ell {p^*}}\\
\end{array}
\right].
\]
By solving the above system of equations for all $\alpha_i$, the
maximum $\hat{\alpha} = \max_i \alpha_i$ can be used to approximate
the true maximum $\max_i \beta_i = \| u^{(m)} \|_\infty$. This
projection can be thought of as querying distinct moments of the
distribution $\pmf_{U^{(m)}}$ that corresponds to some unknown vector
$u^{(m)}$, and then assembling the moments into a model in order to
predict the unknown maximum value in $u^{(m)}$. Of course, when $r$,
the number of terms in our model, is sufficiently large, then
computing $r$ norms of $u^{(m)}$ will result in an exact result, but
it could result in $O(k_m)$ execution time, meaning that our numerical
max-convolution algorithm becomes quadratic; therefore, we must
consider that a small number of distinct moments are queried in order
to estimate the maximum value in $u^{(m)}$. Regardless, the system of
equations above is quite difficult to solve directly via elimination
for even very small values of $r$, because the symbolic expressions
become quite large and because symbolic polynomial roots cannot be
reliably computed when the degree of the polynomial is $>5$. Even in
cases when it can be solved directly, it will be far too inefficient.

For this reason, we solve for the $\alpha_i$ values using an exact,
alternative approach: If we define a polynomial $\gamma(x) =
\left(x-\alpha_1^{p^*}\right)\left(x-\alpha_2^{p^*}\right)\cdots
\left(x-\alpha_r^{p^*}\right)$, then $x \in \{ \alpha_1^{p^*},
\alpha_2^{p^*}, \ldots \alpha_r^{p^*} \} \Leftrightarrow \gamma(x) =
0$. We can expand $\gamma(x) = \gamma_0 + \gamma_1 x + \gamma_2 x^2 +
\cdots + \gamma_r x^r$, and then write

\begin{multline*}
\left[
\begin{array}{ccccc}
\gamma_0 & \gamma_1 & \gamma_2 & \cdots & \gamma_r
\end{array}
\right] \cdot
\left[
\begin{array}{cccc}
\alpha_1^{p^*} & \alpha_2^{p^*} & & \alpha_r^{p^*} \\
\alpha_1^{2 {p^*}} & \alpha_2^{2 {p^*}} & \cdots & \alpha_r^{2 {p^*}} \\
\alpha_1^{3 {p^*}} & \alpha_2^{3 {p^*}} & & \alpha_r^{3 {p^*}} \\
\vdots & \vdots & & \vdots \\
\alpha_1^{\ell {p^*}} & \alpha_2^{\ell {p^*}} & & \alpha_r^{\ell {p^*}} \\
\end{array}
\right] \cdot
\left[
\begin{array}{ccccc}
n_1\\
n_2\\
n_3\\
\vdots\\
n_r\\
\end{array}
\right] = \\
\left[
\begin{array}{ccccc}
\alpha_1^{p^*} \gamma(\alpha_1^{p^*}) & \alpha_2^{p^*} \gamma(\alpha_2^{p^*}) & \alpha_3^{p^*} \gamma(\alpha_3^{p^*}) & \cdots & \alpha_r^{p^*} \gamma(\alpha_r^{p^*})
\end{array}
\right] \cdot
\left[
\begin{array}{ccccc}
n_1\\
n_2\\
n_3\\
\vdots\\
n_r\\
\end{array}
\right] = \\
\left[
\begin{array}{ccccc}
0 & 0 & 0 & \cdots & 0
\end{array}
\right] \cdot 
\left[
\begin{array}{ccccc}
n_1\\
n_2\\
n_3\\
\vdots\\
n_r\\
\end{array}
\right] = 0,
\end{multline*}

which indicates that
\[
\left[
\begin{array}{ccccc}
\gamma_0 & \gamma_1 & \gamma_2 & \cdots & \gamma_r
\end{array}
\right] \cdot
\left[
\begin{array}{c}
\| u^{(m)} \|_{p^*}^{p^*} \\
\| u^{(m)} \|_{2 {p^*}}^{2 {p^*}}\\
\| u^{(m)} \|_{2 {p^*}}^{2 {p^*}}\\
\vdots\\
\| u^{(m)} \|_{\ell {p^*}}^{\ell {p^*}}\\
\end{array}
\right] = 0.
\]

Furthermore, $\gamma(x) = 0, x \neq 0 \Leftrightarrow x^i \gamma(x) = 0, i \in \mathbb{N}$; therefore we can write
\begin{multline*}
\left[
\begin{array}{ccccccccc}
\gamma_0 & \gamma_1 & \gamma_2 & \cdots & \gamma_r & 0 & 0 & \cdots & 0\\
0 & \gamma_0 & \gamma_1 & \gamma_2 & \cdots & \gamma_r & 0 & \cdots & 0\\
0 & 0 & \gamma_0 & \gamma_1 & \gamma_2 & \cdots & \gamma_r & \cdots & 0\\
  &   &     &     & \vdots & \\
0 & 0 & \cdots & 0 & \gamma_0 & \gamma_1 & \gamma_2 & \cdots & \gamma_r\\
\end{array}
\right] \cdot
\left[
\begin{array}{c}
\| u^{(m)} \|_{p^*}^{p^*} \\
\| u^{(m)} \|_{2 {p^*}}^{2 {p^*}}\\
\| u^{(m)} \|_{2 {p^*}}^{2 {p^*}}\\
\vdots\\
\| u^{(m)} \|_{\ell {p^*}}^{\ell {p^*}}\\
\end{array}
\right] = \\
\left[
\begin{array}{ccccc}
\| u^{(m)} \|_{p^*}^{p^*} & \| u^{(m)} \|_{2 {p^*}}^{2 {p^*}} & \| u^{(m)} \|_{3 {p^*}}^{3 {p^*}} & \cdots & \| u^{(m)} \|_{(r+1) {p^*}}^{(r+1) {p^*}} \\
\| u^{(m)} \|_{2 {p^*}}^{2 {p^*}} & \| u^{(m)} \|_{3 {p^*}}^{3 {p^*}} & \| u^{(m)} \|_{4 {p^*}}^{4 {p^*}} & \cdots & \| u^{(m)} \|_{(r+2) {p^*}}^{(r+2) {p^*}} \\
\| u^{(m)} \|_{3 {p^*}}^{3 {p^*}} & \| u^{(m)} \|_{4 {p^*}}^{4 {p^*}} & \| u^{(m)} \|_{5 {p^*}}^{5 {p^*}} & \cdots & \| u^{(m)} \|_{(r+3) {p^*}}^{(r+3) {p^*}} \\
& & \vdots & &\\
\| u^{(m)} \|_{(\ell-r-1) {p^*}}^{(\ell-r-1) {p^*}} & \cdots & \| u^{(m)} \|_{(\ell-2) {p^*}}^{(\ell-2) {p^*}} & \| u^{(m)} \|_{(\ell-1) {p^*}}^{(\ell-1) {p^*}} & \| u^{(m)} \|_{\ell {p^*}}^{\ell {p^*}} \\
\end{array}
\right] \cdot
\left[
\begin{array}{c}
\gamma_0\\
\gamma_1\\
\gamma_2\\
\vdots\\
\gamma_r
\end{array}
\right] = 0.
\end{multline*}

Therefore, 
\[ \left[
\begin{array}{c}
\gamma_0\\
\gamma_1\\
\gamma_2\\
\vdots\\
\gamma_r
\end{array}
\right] \in null\left(
\left[
\begin{array}{ccccc}
\| u^{(m)} \|_{p^*}^{p^*} & \| u^{(m)} \|_{2 {p^*}}^{2 {p^*}} & \| u^{(m)} \|_{3 {p^*}}^{3 {p^*}} & \cdots & \| u^{(m)} \|_{(r+1) {p^*}}^{(r+1) {p^*}} \\
\| u^{(m)} \|_{2 {p^*}}^{2 {p^*}} & \| u^{(m)} \|_{3 {p^*}}^{3 {p^*}} & \| u^{(m)} \|_{4 {p^*}}^{4 {p^*}} & \cdots & \| u^{(m)} \|_{(r+2) {p^*}}^{(r+2) {p^*}} \\
\| u^{(m)} \|_{3 {p^*}}^{3 {p^*}} & \| u^{(m)} \|_{4 {p^*}}^{4 {p^*}} & \| u^{(m)} \|_{5 {p^*}}^{5 {p^*}} & \cdots & \| u^{(m)} \|_{(r+3) {p^*}}^{(r+3) {p^*}} \\
& & \vdots & &\\
\| u^{(m)} \|_{(\ell-r-1) {p^*}}^{(\ell-r-1) {p^*}} & \cdots & \| u^{(m)} \|_{(\ell-2) {p^*}}^{(\ell-2) {p^*}} & \| u^{(m)} \|_{(\ell-1) {p^*}}^{(\ell-1) {p^*}} & \| u^{(m)} \|_{\ell {p^*}}^{\ell {p^*}} \\
\end{array}
\right]
\right).
\]

Because the columns of 
\[ 
\left[
\begin{array}{cccc}
\alpha_1^{p^*} & \alpha_2^{p^*} & & \alpha_r^{p^*} \\
\alpha_1^{2 {p^*}} & \alpha_2^{2 {p^*}} & \cdots & \alpha_r^{2 {p^*}} \\
\alpha_1^{3 {p^*}} & \alpha_2^{3 {p^*}} & & \alpha_r^{3 {p^*}} \\
\vdots & \vdots & & \vdots \\
\alpha_1^{\ell {p^*}} & \alpha_2^{\ell {p^*}} & & \alpha_r^{\ell {p^*}} \\
\end{array}
\right]
\]
must be linearly independent when $\alpha_1, \alpha_2, \ldots$ are
distinct (which is the case by the definition of our multiset
formulation of the norm), then $r=\frac{\ell}{2}$ will determine a
unique solution; thus the null space above is computed from a matrix
with $r+1$ columns and $r$ rows, yielding a single vector for
$(\gamma_0, \gamma_1, \ldots \gamma_r)$. This vector can then be used
to compute the roots of the polynomial $\gamma_0 + \gamma_1 x +
\gamma_2 x^2 + \cdots + \gamma_r x^r$, which will determine the values
$\{\alpha_1^{p^*}, \alpha_2^{p^*}, \ldots \alpha_r^{p^*}\}$, which can
each be taken to the $\frac{1}{{p^*}}$ power to compute $\{\alpha_1,
\alpha_2, \ldots, \alpha_r\}$; the largest of those $\alpha_i$ values
is used as the estimate of the maximum element in $u^{(m)}$. When
$u^{(m)}$ contains at least $r$ distinct values (\emph{i.e.}, $e_m
\geq r$), then the problem will be well-defined; thus, if the roots of
the null space spanning vector are not well-defined, then a smaller
$r$ can be used (and should be able to compute an exact estimate of
the maximum, since $u^{(m)}$ can be projected exactly when $r$ is the
precise number of unique elements found in $u^{(m)}$).

Note that this projection method is valid for any sequence of norms
with even spacing: ${\| u^{(m)} \|}_{p_0 + p^*}^{p^*_0 + p^*}, {\| u^{(m)}
    \|}_{p_0 + 2 p^*}^{p_0 + 2 p^*}, {\| u^{(m)} \|}_{p_0 + 3 p^*}^{p_0 +
        3 p^*}, \ldots {\| u^{(m)} \|}_{p_0 + \ell p^*}^{p_0 +
        \ell p^*}$.

\subsubsection{Closed-Form Projection Method for $r=2$}
In general, the computation of both the null space spanning vector
$(\gamma_0, \gamma_1, \ldots \gamma_r)$ and of machine-precision
approximations for the roots of the polynomial $\gamma_0 + \gamma_1 x
+ \gamma_2 x^2 + \cdots + \gamma_r x^r$ (which can be approximated by
constructing a matrix with that characteristic polynomial and
performing eigendecomposition~\cite{horn2012matrix}) are both in
$O(r^3)$ for each index $m$ in the result; however, by using a small
$r=2$, we can compute a closed form solution of both the null space
spanning vector and of the resulting quadratic roots. This enables
faster exploitation of the curve of norms for estimating the maximum
value of $u^{(m)}$ (although it doesn't achieve the high accuracy
possible with a much larger $r \approx e$). This is equivalent to
approximating $\| u^{(m)} \|_{p^*}^{p^*} \approx h_1 \alpha_1^{p^*} +
h_2 \alpha_2^{p^*}$, where $h_1 + h_2 = k_m = len(u^{(m)})$.

In this case, the single spanning vector of the null space of
\[
\left[
\begin{array}{ccc}
\| u^{(m)} \|_{p^*}^{p^*} & \| u^{(m)} \|_{2 p^*}^{2 p^*} & \| u^{(m)} \|_{3 p^*}^{3 p^*} \\
\| u^{(m)} \|_{2 p^*}^{2 p^*} & \| u^{(m)} \|_{3 p^*}^{3 p^*} & \| u^{(m)} \|_{4 p^*}^{4 p^*}
\end{array}
\right]
\]
will be
\[
\left[
\begin{array}{c}
\gamma_0\\
\gamma_1\\
\gamma_2\\
\end{array}
\right] = 
\left[
\begin{array}{c}
\| u^{(m)} \|_{2 p^*}^{2 p^*} \| u^{(m)} \|_{4 p^*}^{4 p^*} - {\left( \| u^{(m)} \|_{3 p^*}^{3 p^*} \right)}^2\\
\| u^{(m)} \|_{p^*}^{p^*} \| u^{(m)} \|_{4 p^*}^{4 p^*} - \| u^{(m)} \|_{2 p^*}^{2 p^*} \| u^{(m)} \|_{3 p^*}^{3 p^*}\\
\| u^{(m)} \|_{p^*}^{p^*} \| u^{(m)} \|_{3 p^*}^{3 p^*} - {\left( \| u^{(m)} \|_{2 p^*}^{2 p^*} \right)}^2\\
\end{array}
\right]
\]
and thus $\hat{\alpha} \approx \| u^{(m)} \|_\infty$ can be computed
by using the quadratic formula to solve $\gamma_0 + \gamma_1 x +
\gamma_2 x^2 = 0$ for $x$, and computing $\hat{\alpha}$ using the
maximum of those zeros: $\hat{\alpha} = {x_{\max}}^{\frac{1}{p^*}}$. When
the quadratic is not well defined, then this indicates that the number
of unique elements in $u^{(m)}$ is less than 2, and thus cannot be
projected uniquely (\emph{i.e.}, $e_m < r$); in this case, the
closed-form linear solution can be used rather than a closed-form
quadratic solution:
\[
\hat{\alpha} = {\left(\frac{\| u^{(m)} \|_{4 p^*}^{4 p^*}}{\| u^{(m)} \|_{3 p^*}^{3 p^*}}\right)}^\frac{1}{p^*}.
\]
When the closed-form linear solution is not numerically stable (due to
division by a value close to zero), then the $p^*$-norm approximation
can likewise be used.

\subsubsection{Adapted Piecewise Algorithm Using Interleaved $p^*$ Points}

Because the norms must have evenly spaced $p^*$ values in order to use
the projection method described above, the exponential sequence of
$p^*$ values used in the original piecewise algorithm will not contain
four evenly spaced points (which are necessary to solve the quadratic
formulation, \emph{i.e.} $r=2$). One possible solution would be to
take the maximal stable value of $p^*$ for any index (which will be a
power of two found using the original piecewise method), and then also
computing norms (via the FFT, as before) for $p^*-3 \delta,
p^* - 2 \delta, p^* - \delta, p^*$;
however, this will result in a $4\times$ slowdown in the algorithm,
because for every $p^*$-norm computed via FFT before, now four must be
computed. An alternative approach reuses existing values in the $2^i$
sequence of $p^*$: for $p^*$ sufficiently large, then the
exponential sequence is guaranteed to include these stable $p^*$
values: $\frac{p^*}{4}, \frac{p^*}{2},
p^*$. By considering $\frac{3 p^*}{4}$ in $p^*$
candidates, then we can be guaranteed to have four evenly spaced and
stable $p^*$ values. This can be achieved easily by noting that
\[
\frac{3 p^*}{4} = \frac{\frac{p^*}{2} + p^*}{2},
\]
meaning that we can insert all possible necessary $p^*$ values for
evenly spaced sequences of length four by first computing the
exponential sequence of $p^*$ values and then inserting the averages
between every pair of adjacent powers of two (and inserting them in a
way that maintains the sorted order): $1, 2, 4, 8, 16, \ldots$ becomes
$1, 1.5, 2, 3, 4, 6, 8, 12, 16, \ldots$. Thus, if (for some index $m$)
16 is the highest stable $p^*$ that is a power of two (\emph{i.e.},
the $p^*$ value that would be used by the original piecewise
algorithm), then we are guaranteed to use the evenly spaced sequence
$4, 8, 12, 16$. By interleaving the powers of two with the averages
from the following powers of two, we reduce the number of FFTs to
$2\times$ that used by the original piecewise algorithm. For small
values of $r$ (such as the $r=2$ used here), the estimation of the
maximum from each sequence of four norms is in $O(4 k)$, meaning the
total time will still be $k \log(k) \log(\log(k) + 4 k \in O(k \log(k)
\log(\log(k)))$, which is the same as before. Because the spacing in
this formulation is $\frac{p^*}{4}$, and given the maximal
root of the quadratic polynomial $\gamma(x_{\max}) = 0$, then
$\hat{\alpha} = x_{\max}^\frac{4}{p^*}$ (taking the maximal
root $x_{\max}$ to the power $\frac{4}{p^*}$ instead of
$\frac{1}{p^*}$, which had been the spacing used in the description of
the projection method). The null space projection method is shown in
\textbf{Algorithm~\ref{algorithm:piecewiseWithProjection}}.

\begin{algorithm}
  \caption{ {\bf Piecewise numerical max-convolution with projection}, a
    numerical method to estimate the max-convolution of two PMFs or
    nonnegative vectors. This method uses a nullspace projection
    to achieve a closer estimate of the true maximum. Depending
    on the number of stable estimates, linear or quadratic projection
    is used.
    The parameters are two nonnegative vectors
    $L'$ and $R'$ (both scaled so that they have maximal element 1).
    The return value is a numerical estimate of the max-convolution
    $L' ~*_{\max}~R'$.}

  \label{algorithm:piecewiseWithProjection}
  \begin{small}
    \begin{algorithmic}[1]
      \Procedure{numericalMaxConvolvePiecewiseProjectionAffine}{$L'$, $R'$, $p^*$}
      
      \State $\ell_{\max} \gets \argmax_\ell L[\ell]$
      \State $r_{\max} \gets \argmax_r R[r]$
      \State $L' \gets \frac{L}{L[\ell_{\max}]}$
      \State $R' \gets \frac{R}{R[r_{\max}]}$ \Comment{Scale to a proportional problem on $L',R'$}

      \State $allPStar \gets [ 2^{-1}, 2^0, 2^1, \dots, 2+2^{\floor[\big]{\log_2(p^*_{\max})}} ]$
      
      \For{$h \in \{ 0, 1, \ldots len(allPStar)\}$}
      \State $allPStarInterleaved[2i] \gets allPStar[i] $
      \State $allPStarInterleaved[2i+1] \gets 0.5 \times (allPStar[i]+allPStar[i+1]) $
      \EndFor
      
      \For{$i \in \{ 0, 1, \ldots len(allPStar)\}$}
      \State $resForAllPStar[i] \gets$ \texttt{fftNonnegMaxConvolveGivenPStar}($L'$, $R'$, $allPStarInterleaved[i]$)
      \EndFor
      
      \For{$m \in \{ 0, 1, \ldots len(L)+len(R)-1\}$}
      \State $maxStablePStarIndex[m] \gets \max \{ i:~ {\left( resForAllPStar[i][m] \right)}^{\text{allPStarInterleaved[$i$]}} \geq \tau) \}$
      \EndFor
      
      \For{$o \in \{ 0, 1, \ldots len(maxStablePStarIndex)\}$}
      \State $maxStablePStarIndex[o] -= maxStablePStarIndex[o] \% 2$ \Comment{Restrict to powers of $2$}
      \EndFor

      \For{$p \in \{ 0, 1, \ldots len(maxStablePStarIndex)\}$}
      \State $maxP \gets allPStarInterleaved[maxStablePStarIndex[p]]$
      \State $spacing \gets 0.25*maxP$
      \State $est_4 \gets resForAllPStar[maxStablePStarIndex[p]]$
      \State $est_3 \gets resForAllPStar[maxStablePStarIndex[p]-1]$
      \If{$maxStablePStarIndex[p] < 5$} \Comment{Need 5 $p^*$ in sequence to get 4 evenly spaced}
      \State $resForAllPStar[p] \gets$ \texttt{maxLin}$(est_3, est_4)$
      \Else
      \State $est_2 \gets resForAllPStar[maxStablePStarIndex[p]-2]$
      \State $est_1 \gets resForAllPStar[maxStablePStarIndex[p]-4]$ \Comment{Index - 4 is the next evenly spaced point}
      \State $resForAllPStar[p] \gets$ \texttt{maxQuad}$(est_1, est_2, est_3, est_4, spacing)$
      \EndIf
      \EndFor

      \State $result \gets $\texttt{affineCorrect}$(resForAllPStar, maxStablePStarIndex)$

      \State \Return $L[\ell_{\max}] \times R[r_{\max}] \times result$
      \Comment{Undo previous scaling}

      \EndProcedure
    \end{algorithmic}
  \end{small}
\end{algorithm}

\begin{algorithm}
  \caption{ {\bf Linear projection of the maximum}, using previously
    computed values $est_3, est_4$ for two $p^*$ with a difference of
    $spacing$ (estimates given in ascending order of their corresponding
  	$p$'s used).  The naming of the variables
    follows the scheme $est_i = \| u^{(m)} \|_{\frac{i}{4}
      maxP}^{\frac{i}{4} maxP}$.  To prevent numeric instabilities,
    the algorithm checks for division by zero within a tolerance
    $\tau_{\text{Div}} = 10^{-10}$ (again, a conservative estimate of
    the machine precision). The return value is a new estimate of the
    real maximum.}

  \label{algorithm:maxLin}
  \begin{small}
    \begin{algorithmic}[1]
      \Procedure{maxLin}{$est_3$, $est_4$, $spacing$}
      \If{$|est_3| > \tau_{\text{Div}}$}
      \State $result \gets \frac{est_4}{est_3}$
      \Else
      \State $result \gets est_4$
      \EndIf
      \State \Return $result^{(1.0/spacing)}$
      \EndProcedure
    \end{algorithmic}
  \end{small}
\end{algorithm}

\begin{algorithm}
  \caption{ {\bf Quadratic projection of the maximum}, using
  previously computed estimates $est_1,est_2,est_3,est_4$ for four equally spaced $p$
  in steps of $spacing$ (estimates given in ascending order of their corresponding
  $p$'s used).
  The naming of the variables follows the scheme
  $est_i = \| u^{(m)} \|_{\frac{i}{4} maxP}^{\frac{i}{4} maxP}$.
  To prevent numeric
  instabilities, the algorithm checks for division by zero
  within a tolerance $\tau_{\text{Div}} = 10^{-10}$. The return value
  is a new estimate of the real maximum.}

  \label{algorithm:maxQuad}
  \begin{small}
    \begin{algorithmic}[1]
      \Procedure{maxQuad}{$est_1$, $est_2$, $est_3$, $est_4$, $spacing$}
	  \State $\gamma_2 \gets est_1 * est_3 - est_2^2$
	  \State $\gamma_1 \gets est_2 * est_3 - est_1 * est_4$
	  \State $\gamma_0 \gets est_2 * est_4 - est_3^2$
	  
	  \State $preRootValue \gets \gamma_1^2 - 4 * \gamma_2 * \gamma_0$
	  \State $stableQuadratic \gets (\gamma_0 > \tau_{\text{Div}}) ~\&~ (preRootValue >= 0.0)$
	  
	  \If{$stableQuadratic$}
	  \State $result \gets (-\gamma_1 + \sqrt{preRootValue}/(2*\gamma_2)$
	  \Else \Comment{Resort to linear projection}
	  \State $result \gets$ \texttt{maxLin}$(est_3, est_4)$
	  \EndIf
      \State\Return $result^{(1.0/spacing)}$
      \EndProcedure
    \end{algorithmic}
  \end{small}
\end{algorithm}

\subsubsection{Accuracy of the $r=2$ Projection-Based Method}

The full closed-form of the quadratic roots used above (which solve
the projection when $r=2$) will be 

\begin{small}
\begin{align*}
\hat{\alpha} = {}&\max \left( \left(\frac{-\gamma_1 \pm \sqrt{\gamma_1^2 - 4 \gamma_2 \gamma_0}}{2 \gamma_2}\right)^{\frac{1}{p^*}}\right)\\
\begin{split}
 ={}& \max \left( \left(\left(\upp{2} \upp{3} - \upp{1} \upp{4}\right.\right.\right. \\
     & \pm \left.\left. \left((\upp{2} \upp{3} - \upp{1} \upp{4})^2\right.\right.\right. \\
     & - 4 \left.\left.\left. (\upp{1} \upp{3} - \upp{2}^2) (\upp{2} \upp{4} - \upp{3}^2)\right)^{0.5}\right)\right. \\
     & \left.\left. ~\div 2 (\upp{2}^2 - \upp{1} \upp{3})\right)^{\frac{1}{p^*}} \right)
\end{split}\\
\begin{split}
 ={}& \max \left( \left( \| u^{(m)} \|_\infty^{p^*} \left(\vpp{2} \vpp{3} - \vpp{1} \vpp{4}\right.\right.\right. \\
     & \pm \left.\left. \left((\vpp{2} \vpp{3} - \vpp{1} \vpp{4})^2\right.\right.\right. \\
     & - 4 \left.\left.\left. (\vpp{1} \vpp{3} - \vpp{2}^2) (\vpp{2} \vpp{4} - \vpp{3}^2)\right)^{0.5}\right)\right. \\
     & \left.\left. ~\div 2 (\vpp{2}^2 - \vpp{1} \vpp{3})\right)^{\frac{1}{p^*}} \right)
\end{split}\\
\begin{split}
 ={}& \| u^{(m)} \|_\infty \max \left( \left(\left(\vpp{2} \vpp{3} - \vpp{1} \vpp{4}\right.\right.\right. \\
     & \pm \left.\left. \left((\vpp{2} \vpp{3} - \vpp{1} \vpp{4})^2\right.\right.\right. \\
     & - 4 \left.\left.\left. (\vpp{1} \vpp{3} - \vpp{2}^2) (\vpp{2} \vpp{4} - \vpp{3}^2)\right)^{0.5}\right)\right. \\
     & \left.\left. ~\div 2 (\vpp{2}^2 - \vpp{1} \vpp{3})\right)^{\frac{1}{p^*}} \right)
\end{split}\\
\end{align*}
\end{small}
where $p^* = \frac{maxP}{4}$ in the pseudocode (\emph{i.e.}, the
maximum numerically stable $p^*$ used by the piecewise algorithm at
that index). Note that $\| u^{(m)} \|_\infty^{p^*}$ can be factored
out because the exponents in every term in the numerator will be
$5p^*$ (\emph{i.e.}, $10 p^*$ in the square root). Similarly the terms
in the denominator each contain $\| u^{(m)}
\|_\infty^{4p^*}$. Factoring out the maximum value is then the same as
operating on scaled vectors $v$ (instead of $u$) with the maximum
entry being $1$, and at least one element of value $1$.

Furthermore, the denominator $2 \gamma_2 \geq 0$; even though the
terms summed to compute $\gamma_2$ are not exclusively nonnegative,
symmetry can be used to demonstrate that every negative term is
outweighed by a unique corresponding term:
\begin{eqnarray*}
\gamma_2 & = & \| v^{(m)} \|_{1} \| v^{(m)} \|_{3}^{3} - {\left( \| u^{(m)} \|_{2 p^*}^{2 p^*} \right)}^2 \\
& = & \left( \sum_i {v^{(m)}_i} \right) \left( \sum_i {v^{(m)}_i}^3 \right) - {\left( \sum_i {v^{(m)}_i}^2 \right)}^2 \\
& = & \sum_{i,j} {v^{(m)}_i} {v^{(m)}_j}^3 - \sum_{i,j} {v^{(m)}_i}^2 {v^{(m)}_j}^2 \\
& = & \sum_{i,j} {v^{(m)}_i} {v^{(m)}_j}^2 \left( {v^{(m)}_j} - {v^{(m)}_i} \right) \\
& = & \sum_{i} {v^{(m)}_i} {v^{(m)}_i}^2 \left( {v^{(m)}_i} - {v^{(m)}_i} \right) + \sum_{i<j} {v^{(m)}_i} {v^{(m)}_j}^2 \left( {v^{(m)}_j} - {v^{(m)}_i} \right) + {v^{(m)}_j} {v^{(m)}_i}^2 \left( {v^{(m)}_i} - {v^{(m)}_j} \right) \\
& = & 0 + \sum_{i<j} {v^{(m)}_i} {v^{(m)}_j} \left( {v^{(m)}_j} - {v^{(m)}_i} \right) \left( {v^{(m)}_j} - {v^{(m)}_i} \right)\\
& = & \sum_{i<j} {v^{(m)}_i} {v^{(m)}_j} {\left( {v^{(m)}_j} - {v^{(m)}_i} \right)}^2\\
& \geq & 0.\\
\end{eqnarray*}
Thus, for well-defined problems (\emph{i.e.}, when $\gamma_2 \neq 0$),
the denominator $2 \gamma_2 > 0$, and therefore, the maximum root of
the quadratic polynomial will correspond to the term that adds (rather
than subtracts) the square root term:
\begin{align*}
\begin{split}
\hat{\alpha} 
 ={}& \| u^{(m)} \|_\infty \max \left( \left(\left(\vpp{2} \vpp{3} - \vpp{1} \vpp{4}\right.\right.\right. \\
     & + \left.\left. \left((\vpp{2} \vpp{3} - \vpp{1} \vpp{4})^2\right.\right.\right. \\
     & - 4 \left.\left.\left. (\vpp{1} \vpp{3} - \vpp{2}^2) (\vpp{2} \vpp{4} - \vpp{3}^2)\right)^{0.5}\right)\right. \\
     & \left.\left. ~\div 2 (\vpp{2}^2 - \vpp{1} \vpp{3})\right)^{\frac{1}{p^*}} \right).
\end{split}
\end{align*}

The relative absolute error is defined as $| \frac{\hat{\alpha} - \|
  u^{(m)} \|_\infty}{\| u^{(m)} \|_\infty} | = |
\frac{\hat{\alpha}}{\| u^{(m)} \|_\infty} - 1|$; therefore, a bound on
the relative error of the projection method can be established by
bounding
\begin{align*}
s(p^*, k_m) ={}& \frac{\hat{\alpha}}{\| u^{(m)} \|_\infty} \\
\begin{split}
 ={}& \left(\left(\vpp{2} \vpp{3} - \vpp{1} \vpp{4}\right.\right. \\
     & + \left.\left. \left((\vpp{2} \vpp{3} - \vpp{1} \vpp{4})^2\right.\right.\right. \\
     & - 4 \left.\left.\left. (\vpp{1} \vpp{3} - \vpp{2}^2) (\vpp{2} \vpp{4} - \vpp{3}^2)\right)^{0.5}\right)\right. \\
     & \left. ~\div 2 (\vpp{2}^2 - \vpp{1} \vpp{3})\right)^{\frac{1}{p^*}}.
\end{split}
\end{align*}
where the length of the $u^{(m)}$ (respectively $v^{(m)}$) is $k_m$. Using this reformulation,
$s = 1$ indicates a zero-error approximation. This can be rewritten to
bound its value before taking to the power $\frac{1}{p^*}$:
\[
s(p^*, k_m) = {t(p^*, k_m)}^{\frac{1}{p^*}}
\]
where 
\begin{align*}
\begin{split}
t(p^*, k_m)
={}& \left(\vpp{2} \vpp{3} - \vpp{1} \vpp{4}\right.\\
& + \left.\left. \left((\vpp{2} \vpp{3} - \vpp{1} \vpp{4})^2\right.\right.\right. \\
& - 4 \left.\left.\left. (\vpp{1} \vpp{3} - \vpp{2}^2) (\vpp{2} \vpp{4} - \vpp{3}^2)\right)^{0.5}\right)\right. \\
& ~\div 2 (\vpp{2}^2 - \vpp{1} \vpp{3}).
\end{split}
\end{align*}


The extreme values of $t(p^*, k_m)$ can be found by minimizing and
maximizing over the possible values of $v^{(m)} \in V = \{ v:
{[0,1]}^{k_m} : \exists i, v_i = 1, \exists j, v_j \in (0,1) \}$. The
final constraint on $v_j$ in (0,1) is because any $v$ containing only
one unique value (which must be $1$ in this case since dividing by the
maximum element in $u^{(m)}$ to compute $v^{(m)}$ has divided the
value at that index by itself ($\exists i, v_i = 1$) will lead to
instabilities. When values in
$v$ are identical to one another, using $r=1$ yields an exact
solution, and thus solving with $r=2$ is not well-defined because
$\gamma_2 = 0$. Because all elements ${v^{(m)}}^{p^*} \in [0,1]$ and
$p^* \geq 1$, we can perform a change of variables $v^{(m)}_i = {v^{(m)}}_i^{p^*}$, thereby eliminating references to $p^*$:
\begin{align*}
\begin{split}
t(k_m) \geq{}& \min_{ v \in \mathbb{R}^{k_m} : \exists i, v_i = 1,
  \exists j, v_j \in (0,1)}
\left(\vp{2} \vp{3} - \vp{1} \vp{4}\right.\\
& + \left.\left. \left((\vp{2} \vp{3} - \vp{1} \vp{4})^2\right.\right.\right. \\
& - 4 \left.\left.\left. (\vp{1} \vp{3} - \vp{2}^2) (\vp{2} \vp{4} - \vp{3}^2)\right)^{0.5}\right)\right. \\
& ~\div 2 (\vp{2}^2 - \vp{1} \vp{3}).
\end{split}
\end{align*}
\begin{align*}
\begin{split}
t(k_m) \leq{}& \max_{ v \in \mathbb{R}^{k_m} : \exists i, v_i = 1,
  \exists j, v_j \in (0,1)}
\left(\vp{2} \vp{3} - \vp{1} \vp{4}\right.\\
& + \left.\left. \left((\vp{2} \vp{3} - \vp{1} \vp{4})^2\right.\right.\right. \\
& - 4 \left.\left.\left. (\vp{1} \vp{3} - \vp{2}^2) (\vp{2} \vp{4} - \vp{3}^2)\right)^{0.5}\right)\right. \\
& ~\div 2 (\vp{2}^2 - \vp{1} \vp{3}).
\end{split}
\end{align*}

\begin{table}
\centering
\begin{tabular}{r|ccccc}
\toprule
$k_m$ & 3 & 4 & 5 & 6 & 7 \\
\midrule \\
Minimum & 0.935537 & 0.902161 & 0.895671 & 0.880487 & 0.85343 \\
Maximum & 1 & 1 & 1 & 1 & 1\\
\bottomrule
\end{tabular}
\caption{{\bf Exact bounds of $t(k_m)$ for short vectors of length
    $k_m$.}  This table shows the results of numerical minimization
  techniques performed on the symbolic closed-form of $t(k_m)$ in
  Mathematica (using \texttt{NMinimize}).  All $k_m - 1$ entries
  (excluding the first that was set to $1.0$) were left symbolic and
  constrained to $[0,1]$, with restriction that the denominator of
  $t(k_m)$ was nonzero.}
\label{table:boundsmathematica}
\end{table}
For small vector lengths, the exact bounds of $t(k_m)$ are shown in
\textbf{Table~\ref{table:boundsmathematica}}. Notice that the upper bound is
fixed, but the lower bound grows monotonically smaller as $k_m$, the
length of the vector considered, increases. For larger vectors,
Mathematica does not find optima in a matter of hours, and for
arbitrary-length vectors, the Karush-Kuhn-Tucker criteria do not
easily yield minima or maxima; however, we do observe that all maxima
are achieved by vectors that are permutations (order does not
influence the result) of $v = (1, 1, \ldots 1, b, b, \ldots b)$
(again, when only two unique values are found in $v$, the
approximation is exact and thus $\frac{\hat{\alpha}}{\| u^{(m)}
  \|_\infty} = 1$). Likewise, the minima are achieved by permutations
of $v = (1, a, b, b, \ldots b)$. For this reason, we perform further
empirical estimation of the bound by randomly sampling vectors of the
form $(1, v_2, v_3, \ldots v_{k_m})$ with $k_m-1$ degrees of freedom
(d.o.f.)  and sampling vectors of the form $v = (1, a, b, b, \ldots
b)$ (with $2$ d.o.f.), whose extrema are shown in
\textbf{Table~\ref{table:boundspysim}}.

\begin{table}
\begin{tabular}{r|ccccc}
\toprule
$k_m$ & 4 & 64 & 1024 \\
\midrule \\
Minimum ($k_m-1$ d.o.f.) & 0.90221268 & 0.74942834 & 0.81858283 \\
Maximum ($k_m-1$ d.o.f.) & 0.99999986 & 0.92482416 & 0.86795636 \\
\midrule \\
Minimum (vectors of form $(1, a, b, \ldots b)$, $2$ d.o.f.) & 0.90216688 & 0.75455478 &	0.71695386 \\
Maximum (vectors of form $(1, a, b, \ldots b)$, $2$ d.o.f.) & 1.00000000 & 1.0000000 & 1.00000000\\
\bottomrule
\end{tabular}
\caption{{\bf Bounds via random sampling for vectors different in size
    and type.} This table shows the minimal and maximal values
  resulting from the evaluation of $t(k_m)$ on $10^5$ randomly
  generated vectors (uniform distribution in $[0,1]$). The first part
  shows the result for vectors of potentially unconstrained
  composition, besides one (w.l.o.g. the first) being set to
  $1.0$. The values in the second half were obtained based on vectors
  of (supposedly) worst-case composition (i.e. of form $(1, a, b,
  \ldots b)$).}
\label{table:boundspysim}
\end{table}
At length $64$ we see that due to an extreme value scenario, an
unconstrained vector scores slightly lower than a vector holding the
worst-case pattern $(1, a, b, \ldots b)$, because both forms of
sampling approach the true lower bound, but one of the unconstrained
$k_m-2$ d.o.f. is slightly closer.

From these results, we conjecture that $t$ is bounded above $\leq 1$
(this is achievable at any length $k_m$ by letting $v$ contain exactly
two distinct values). In this manner, we achieve our predicted upper
bound of $1$ regardless of the length $k_m$. Likewise, we conjecture
that at any $k_m$ (not simply the lengths investigated, where this
principle is true), the lower bound is given by vectors of the form
$(1, a, b, b, \ldots b)$. Qualitatively, this conjecture stems from
the fact that since the estimate is perfect when $v$ contains exactly
two distinct elements, then the worst-case lower bound when $v$
contains three distinct values will concentrate the points at some
value far from the other two distinct values. When four distinct
values are permitted, then we conjecture that the optimal choice (for
minimizing $t$) of the fourth value will equal the choice for the
third distinct value, since that was already determined to be the best point
for deceiving the quadratic approximation. From this conjecture, we
can then use the fact that the bounds should only grow more extreme as
$k_m$ increases, since $\mathbb{R}^1 \subset \mathbb{R}^2 \subset
\cdots$ (\emph{i.e.} lower-dimensional solutions can always be reached
in a higher dimension by setting some of the values to $0$). Thus the
minimum for any possible vector should be conservatively bounded below
on vectors of the form $(1, a, b, b, \ldots b)$ and is achieved by letting
$k_m$ approach $\infty$:
\begin{multline*}
\lim_{k_m \rightarrow \infty} t(k_m) =\\
\frac{a^4 b-a^3 b^2-a^2 b^3+\sqrt{b^2 \left(-a^4+3 a^3 b-3 a^2 b^2+a b^3+(b-1)^3\right)^2}+a b^4+b^4-b^3-b^2+b}{2 b \left(a^3-2 a^2 b+a b^2+(b-1)^2\right)}.
\end{multline*}
The minimum value of this expression over all $a \in [0,1], b \in
[0,1]$ is $0.704$ (computed again with Mathematica). Overall, assuming
our conjecture regarding the forms of the vectors achieving the minima
and maxima, then it follows that $t \in (0.7, 1]$, and the worst-case
  relative error at the $p^*_{\max}$ contour will be bounded
\[
| t^{\frac{1}{p^*_{\max}}} - 1 | <
1 - {0.7}^{\frac{4}{p^*_{\max}}}.
\]
The steeper decrease in relative error as $p^*$ is increased means
that the same procedure can be used to achieve an absolute error
bound:
\[
\hat{\alpha} - \| u^{(m)} \|_\infty < \tau^{\frac{1}{2 p^*}} \left( 1
- {0.7}^{\frac{4}{p^*}} \right),
\]
which achieves a unique maximum at
\[
p^*_{mode} = \frac{1.4267*\log(\tau) - 4.07094}{ \left(\log(\tau) - 2.8534\right)\left(\log(1-\frac{2.8534}{\log(\tau)}) \right) }
\approx 14.52.
\]
As before, the worst-case absolute error of the unscaled problem will
be found by simply scaling the absolute error at $p^*_{mode}$:
\[
\max_\ell L[\ell]~ \max_r R[r]~ \tau^{\frac{1}{2 p^*_{mode}}} \left( 1 -
    {0.7}^{\frac{4}{p^*_{mode}}} \right).
\]
Because $p^*_{mode}$ (the value of $p^*$ producing the worst-case
absolute error) for the null space projection method it is invariant
to the length of the list $k$ (enabling us to compute a numeric
value), and because its numeric value is so small, even a fairly small
choice of $p^*_{\max}$ will suffice (now $p^*_{\max} \in O(1)$ rather
than in $O(\log(k))$ as it was with the original piecewise
method). For example, the approximation of the Viterbi path to infer
the unemployment data is slightly superior with the null space
projection method, even when $p^*_{\max}=64$ is used (in contrast to
the $p^*_{\max}=8192$ used in the Figure~\ref{figure:viterbi}). The
null space projection method required $136.6$ seconds (slightly faster
than the $141.4$ seconds required by the original piecewise method).

The one caveat of this worst-case absolute error bound is that it
presumes at least four evenly spaced, stable $p^*$ can be found (which
may not be the case by choosing $p^*$ from the sequence $2^i$ in cases
when $\| u^{(m)} \|_\infty \approx 0$); however, assuming standard
fast convolution can be performed (a reasonable assumption given it is
one of the essential numeric algorithms), then four evenly spaced
$p^*$ values could be chosen very close to $1$; therefore, these
values of $p^*$ could be added to the sequence so that the algorithm
is slightly slower, but essentially always yield this worst-case
absolute error bound.

\begin{figure*}
  \centering
  \includegraphics[width=5in]{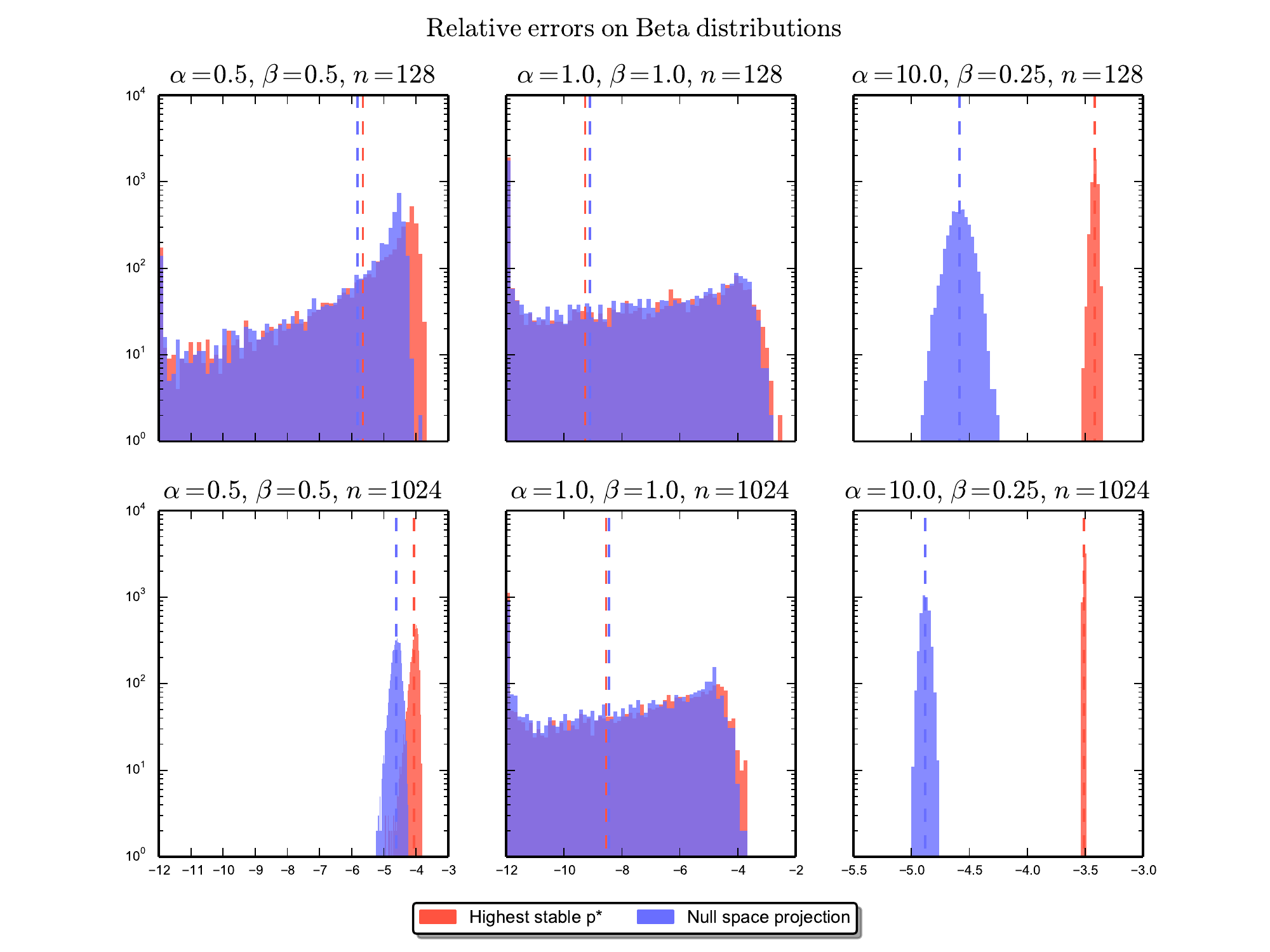}
  \caption{{\bf Relative errors on random vectors with and without
      null space projection.} For the two approximation methods (using
    the highest stable $p^*$-norm with the heuristically chosen
    $p^*_{\max}$ or using the null space projection with
    $p^*_{\max}=64$), vectors of different lengths are sampled
    ($2^{12}$ repetitions) from a variety of Beta distributions. The
    settings for the parameters ($\alpha,\beta$) of the Beta
    distribution that were used, as well as the lengths of the
    generated vectors are shown in the titles of the subplots: $\alpha
    = 0.5, \beta = 0.5$ (bimodal with modes near zero and one);
    $\alpha = 0.1, \beta = 0.1$ (uniform distribution); $\alpha = 10,
    \beta = 0.25$ (with a strong mode near one). The red area
    depicts the frequencies (y-axis) of the different magnitudes of
    (relative) error (x-axis) when using the highest stable $p^*$-norm
    is used as an approximation of the Chebyshev norm
    ($p=\infty$). The blue area shows the errors with the
    method that performs a projection (either quadratic or linear
    depending on how many numerically stable $p^*$ are available) to
    estimate the Chebyshev norm.
  \label{figure:noExtrapolVsExtrapol}}
\end{figure*}

\begin{figure*}
  \centering
  \includegraphics[width=5in]{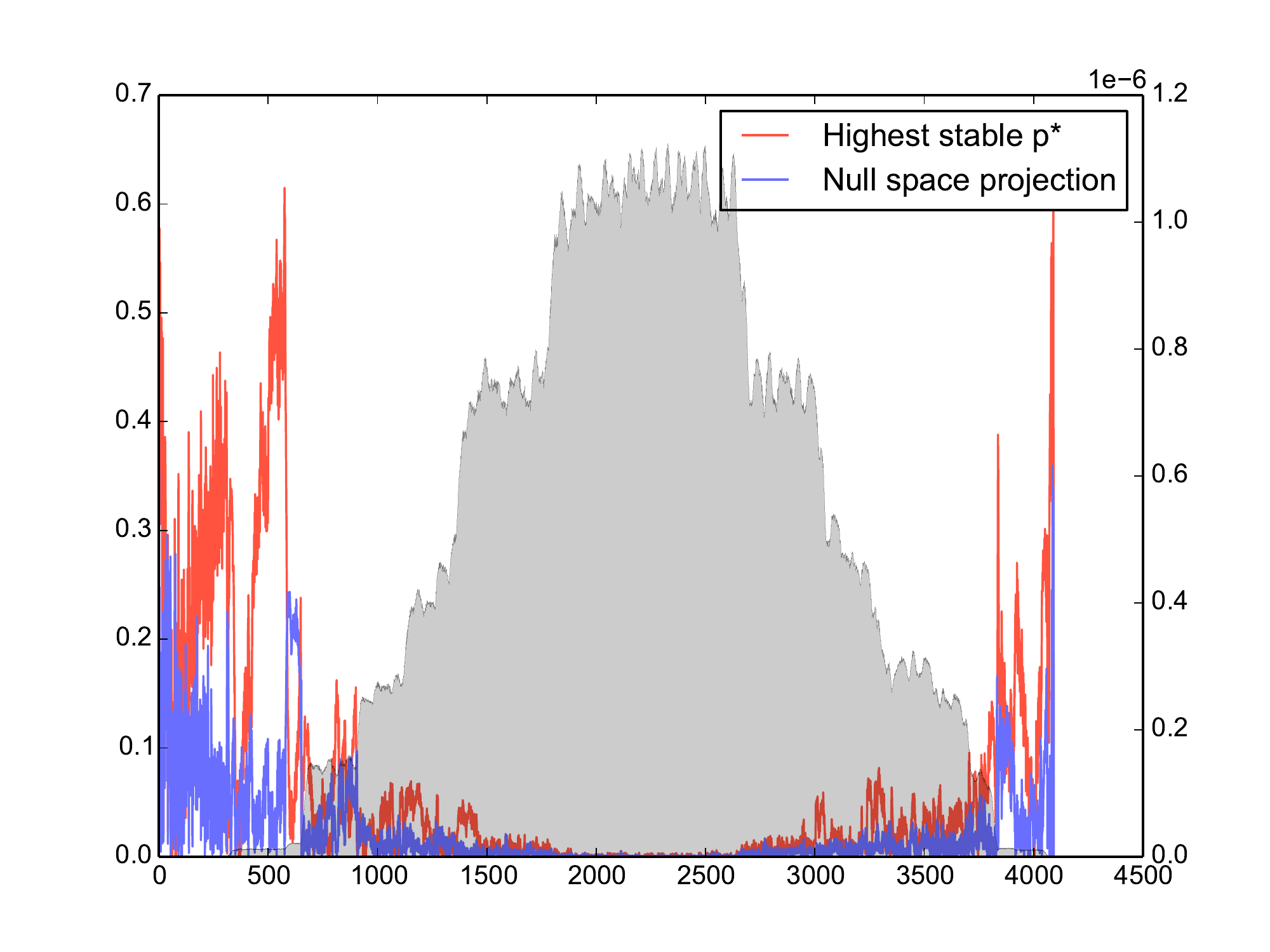}
  \caption{{\bf Relative errors on large max-convolution with and
      without null space projection.} Max-convolution between two
    randomly generated vectors (both uniform vectors convolved with
    narrow Gaussians with uniform noise added afterward), performed
    with the highest stable $p^*$-norm (using the heuristic choice of
    $p^*_{\max}$ for problems of this size) and with null space
    projection (using $p^*_{\max}=64$). The left y-axis shows the
    relative error at index $m$.  Associated with that, you can see
    the red and blue curve depicting the errors from the two
    different methods: Red describes the max-norm estimation using
    only the highest stable $p^*$ while purple was generated using
    quadratic projection at the four highest stable $p^*$ values (when
    at least four evenly spaced values are numerically stable) and
    linear projection at the two highest stable $p^*$ values (when
    only two $p^*$ are numerically stable). The results of both
    approaches are corrected with the affine transformation method
    proposed in this manuscript. In the background the gray shaded
    curve shows the exact result of the max-convolution at every index
    (to be used with the second y-axis on the right).
  \label{figure:affineVsAffineExtrapol}}
\end{figure*}

In practice, we can demonstrate that the null space projection method
is very accurate. First we show the impact of using the quadratic
(\emph{i.e.}, $r=2$) projection method on unscaled single $u^{(m)}$
vectors. The projection method was tested on vectors of different
lengths drawn from different types of Beta distributions and are
compared with the results of the $p$-norms with the highest stable $p$
(Figure~\ref{figure:noExtrapolVsExtrapol}). The relative errors
between the original piecewise method and the null space projection
method are compared using a max-convolution on two randomly created
input PMFs of lengths 1024
(Figure~\ref{figure:affineVsAffineExtrapol}). Note that the null space
projection can also be paired with affine scaling on the back end,
just as the original piecewise method can be. In practice, the null
space projection increases the accuracy demonstrably on a variety of
different problems, although the original piecewise method also
performs well.

Although the worst-case runtime of the null space projection method is
roughly $2\times$ that of the original piecewise method, the error
bound no longer depends on the length of the result $k$. Thus, for a
given relative error bound on the top contour (\emph{i.e.}, the
equivalent of the derivation of $p^*_{\max}$ in the original piecewise
algorithm), the value of $p^*_{\max}$ is fixed and no longer $\in
O(\log(k))$. For example, achieving a $0.5\%$ relative error in the top
contour would require
\[
1-{0.7}^{\frac{4}{p^*_{\max}}} \leq 0.005
\rightarrow
{p^*_{\max}} \geq 4 \frac{\log(0.7)}{\log(0.995)} \approx 284.62,
\]
meaning that choosing $p^*_{\max} = 512$ would achieve a very high
accuracy, but while only performing $2\times 9$ FFTs. For very large
vectors, this will not be substantially more expensive than the
original piecewise algorithm, which uses a higher value of
$p^*_{\max}$ (in this case, $p^*_{\max} = \log_{1.005}(k)$, which
continues to grow as $k$ does) to keep the error lower in practice. As
a result, the runtime of the null space projection approximation is
$\in O(k \log(k))$ rather than $O(k \log(k) \log(\log(k)))$, despite
the similar runtime in practice to the original piecewise method (the
null space projection method uses $2\times$ as many FFTs performed per
$p^*$ value, but requires slightly fewer $p^*$ values).

\subsubsection{Practical runtime comparison}

\begin{table}
\centering
\begin{tabular}{r|ccccccc}
\toprule
$k$ 									& $2^{6}$ & $2^{7}$ & $2^{8}$ & $2^{9}$ & $2^{10}$ & $2^{11}$ & $2^{12}$ \\
\midrule \\
Naive 								& \textbf{0.0142} & 0.0530 & 0.192 & 0.767 & 3.03 & 12.1 & 48.2 \\
Naive (vectorized) 					& 0.0175 & 0.0381 & 0.0908 & 0.251 & 0.790 & 2.75 & 10.1 \\
FILL1 (\citet{Bussieck1994})			& 0.0866 & 1.09 & 7.21 & 19.4 & 457 & --- & --- \\
Max. stable $p^*$, affine corrected	& 0.0277 & 0.0353 & 0.0533 & 0.0848 & 0.149 & 0.274 & 0.537\\
Projection, affine corrected 		& 0.0236 & \textbf{0.0307} & \textbf{0.0467} & \textbf{0.0760} & \textbf{0.137} & \textbf{0.258} & \textbf{0.520}\\
\bottomrule
\end{tabular}
\caption{{\bf Runtimes of different methods for max-convolution on
    uniform vectors of length $k$.}  The runtimes were gathered using
  the \texttt{timeit} package in Python. They include all
  preprocessing steps necessary for the algorithm (\emph{e.g.} sorting
  prior to the FILL1 approach).  The values are total runtimes (in
  seconds) to run $5$ repetitions on different, randomly generated
  vectors. FILL1 was not run on larger problems, because it ran
  substantially longer than the non-vectorized naive approach. On the
  two approximation methods presented in this manuscript, the highest
  stable $p^*$-norm approximation was run with the heuristically
  chosen $p^*_{\max}$ for problems of the appropriate size and the
  null space projection was run with $p^*_{\max}=64$.}
\label{table:runtimesInclBussieck}
\end{table}

To compare the actual runtimes of the final algorithm developed in
this manuscript with a naive max-convolution and a previously proposed
method from \citet{Bussieck1994}, all methods were run on vectors of
different random (uniform in $[0,1]$) composition and length
($k$). The first and second input vector were generated seperately but
are always of same length. \textbf{Table~\ref{table:runtimesInclBussieck}}
shows the result of this experiment. All methods were implemented in
Python, using {\tt numpy} where applicable (\emph{e.g.} to
vectorize). A non-vectorized version of naive max-convolution was
included to estimate the effects of vectorization. The approach from
Bussieck et al. ran as a reimplementation based on the pseudocode in
their manuscript. From their variants of proposed methods, FILL1 was
chosen because of its use in their corresponding benchmark and its
recommendation by the authors for having a lower runtime constant in
practice compared to other methods they proposed. The method is based
on sorting the input vectors and traversing the (implicitly) resulting
partially ordered matrix of products in a way that not all entries
need to be evaluated, while only keeping track of the so-called cover
of maximal elements. FILL1 already includes some more sophisticated
checks to keep the cover small and thereby reducing the overhead per
iteration. Unfortunately, although we observed that the FILL1 method
requires between $O(n \log(n))$ and $O(n^2)$ iterations in practice,
this per-iteration overhead results in a worst-case cost of $\log(n)$
per iteration, yielding an overall runtime in practice between $O(n
\log(n) \log(n))$ and $O(n^2 \log(n))$. As the authors state, this
overhead is due to the expense of storing the cover, which can be
implemented \emph{e.g.} using a binary heap (recommended by
the authors and used in this reimplementation). Additionally, due to
the fairly sophisticated datastructures needed for this algorithm it
had a higher runtime constant than the other methods presented here,
and furthermore we saw no means to vectorize it to improve the
efficiency.  For this reason, it is not truly fair to compare the raw
runtimes to the other vectorized algorithms (and it is not likely that
this Python reimplementation is as efficient as the original version,
which \citet{Bussieck1994} implemented in ANSI-C); however, comparing
a non-vectorized implementation of the naive $O(n^2)$ approach with
its vectorized counterpart gives an estimated $\approx 5\times$
speedup from vectorization, suggesting that it is not substantially
faster than the naive approach on these problems (it should be noted
that whereas the methods presented here have tight runtime bounds but
produce approximate results, the FILL1 algorithm is exact, but its
runtime depends on the data processed). During investigation of these
runtimes, we found that on the given problems, the proposed average
case of $O(n \log(n))$ iterations was rarely reached. A reason might
be an unrecognized violation of the assumptions of the theory behind
this theoretical average runtime in how the input vectors were
generated.

In contrast to the exact method from \citet{Bussieck1994}, the herein
proposed approximate procedure are faster whenever the input vectors
are at least $128$ elements long (shorter vectors are most efficiently
processed with the naive approach). The null space projection method
is the fastest method presented here (because it can use a lower
$p^*_{\max}$), although the higher density of $p^*$ values it uses
(and thus, additional FFTs) make the runtimes nearly identical for
both approximation methods.

\section{Discussion}\label{conclusions}

Both piecewise numerical max-convolution methods are highly accurate
in practice and achieve a substantial speedup over both the naive
approach and the approach proposed by \citet{Bussieck1994}. This is
particularly true for large problems: For the original piecewise
method presented here, the $\log_2(\log_{1+\tau^\frac{1}{4}}(k))$
multiplier may never be small, but it grows so slowly with $k$ that it
will be $<18$ even when $k$ is on the same order of magnitude as the
number of particles in the observable universe. This means that, for
all practical purposes, the method behaves asymptotically as a
slightly slower $O(k \log_2(k))$ method, which means the speedup
relative to the naive method becomes more pronounced as $k$ becomes
large. For the second method presented (the null space projection),
the runtime for a given relative error bound will be in $O(k
\log_2(k))$. In practice, both methods have similar runtime on large
problems.

The basic motivation of the first approach described-- \emph{i.e.},
the idea of approximating the Chebyshev norm with the largest
$p^*$-norm that can be computed accurately, and then convolving
according to this norm using FFT-- also suggests further possible
avenues of research. For instance, it may be possible to compute a
single FFT (rather than an FFT at each of several contours) on a more
precise implementation of complex numbers. Such an implementation of
complex values could store not only the real and imaginary components,
but also other much smaller real and imaginary components that have
been accumulated through $+$ operations, even those which have small
enough magnitudes that they are dwarfed by other summands. With such
an approach it would be possible to numerically approximate the
max-convolution result in the same overall runtime as long as only a
bounded ``history'' of such summands was recorded (\emph{i.e.}, if the
top few magnitude summands---whether that be the top 7 or the top
$\log_2(\log_{1+\tau^{\frac{1}{4}}}(k))$---was stored and operated
on). In a similar vein, it would be interesting to investigate the
utility of complex values that use rational numbers (rather than
fixed-precision floating point values), which will be highly precise,
but will increase in precision (and therefore, computational
complexity of each arithmetic operation) as the dynamic range between
the smallest and largest nonzero values in $L$ and $R$ increases
(because taking $L'$ to a large power $p^*$ may produce a very small
value). Other simpler improvements could include optimizing the error
vs. runtime trade-off between the log-base of the contour search: the
method currently searches $\log_2(p^*_{\max})$ contours, but a smaller
or larger log-base could be used in order to optimize the trade-off
between error and runtime.

It is likely that the best trade-off will occur by performing the fast
$p^*$-norm convolution with a number type that sums values over vast
dynamic ranges by appending them in a short (\emph{i.e.}, bounded or
constant size) list or tree and sums values within the same dynamic
range by querying the list or tree and then summing in at the
appropriate magnitude. This is reminiscent of the fast multipole
algorithm~\citep{rokhlin1985rapid}. This would permit the method to use
a single large $p^*$ rather than a piecewise approach, by
moving the complexity into operations on a single number rather than
by performing multiple FFTs with simple floating-point numbers.

The basic motivation of the second approach described-- \emph{i.e.},
using the \emph{sequence} of $p^*$-norms (each computed via FFT) to
estimate the maximum value-- generalizes the $p^*$-norm fast
convolution numerical approach into an interesting theoretical problem
in its own right: given an oracle that delivers a small number of
norms (the number of norms retrieved must be $c \in o(k)$ to
significantly outperform the naive quadratic approach) about each
vector $u^{(m)}$, amalgamate these norms in an efficient manner to
estimate the maximum value in each $u^{(m)}$. This method may be
applicable to other problems, such as databases where the maximum
values of some combinatorial operation (in this case the \emph{maximum
  a posteriori} distribution of the sum of two random variables $X+Y$)
is desired but where caching all possible queries and their maxima
would be time or space prohibitive. In a manner reminiscent of how we
employ FFT, it may be possible to retrieve moments of the result of
some combinatoric combination between distributions on the fly, and
then use these moments to approximate true maximum (or, in general,
other sought quantities describing the distribution of interest).

In practice, the worst-case relative error of our quadratic
approximation is quite low. For example, when $p^* = 8$ is stable,
then the relative error is less than $2.3\%$, regardless of the
lengths of the vectors being max-convolved. In contrast, the
worst-case relative error using the original piecewise method would be
$\leq k^\frac{1}{16}-1$, where $k$ is the length of the
max-convolution result (when $n=1024$, the relative error of the
original piecewise method would be $\approx 54\%$). 

Of course, the use of the null space projection method is predicated
on the existence of at least four sequential $p^*$ points, but it
would be possible to use finer spacing between $p^*$ values
(\emph{e.g.}, $p^* \in (1, 1.01, 1.02, 1.03)$ to guarantee that this
will essentially be the case as long as FFT (\emph{i.e.} $p^*=1$) is
stable. But more generally, the problem of estimating extrema from
$p^*$-norms (or, equivalently, from the $p^*$-th roots of the $p^*$-th
moments of a distribution with bounded support), will undoubtedly
permit many more possible approaches that we have not yet
considered. One that would be compelling is to relate the Fourier
transform of the sequential moments to the maximum value in the
distribution; such an approach could permit all stable $p^*$ at any
index $m$ to be used to efficiently approximate the maximum value (by
computing the FFT of the sequence of norms). Such new adaptations of
the method could permit low worst-case error without any noticable
runtime increase.

\subsection{Multidimensional Numerical Max-Convolution}

The fast numerical piecewise method for max-convolution (and the
affine piecewise modification) are both applicable to matrices as well
as vectors (and, most generally, to tensors of any dimension). This is
because the $p^*$-norm (as well as the derived error bounds as an
approximation of the Chebyshev norm) can likewise approximate the
maximum element in the tensor $u^{(m_1, m_2, \ldots)}$ generated to
find the max-convolution result at index $m_1, m_2, \ldots$ of a
multidimensional problem, because the sum
\[ \sum_{i_1, i_2, \ldots} {\left( u^{(m_1, m_2, \ldots)}_{i_1, i_2, \ldots} \right)}^{p^*} \]
computed by convolution corresponds to the Frobenius norm (\emph{i.e.}
the ``entrywise norm'') of the tensor, and after taking the result of
the sum to the power $\frac{1}{p^*}$, will converge to the maximum
value in the tensor (if $p^*$ is large enough).

This means that the fast numerical approximation, including the affine
piecewise modification, can be used without modification by invoking
standard multidimensional convolution (\emph{i.e.}, $*$). Matrix (and,
in general, tensor) convolution is likewise possible for any dimension
via the row-column algorithm, which transforms the FFT of a matrix
into sequential FFTs on each row and column. The accompanying Python
code demonstrates the fast numerical max-convolution method
on matrices, and the code can be run on tensors of any dimension
(without requiring any modification).

The speedup of FFT tensor convolution (relative to naive convolution)
becomes considerably higher as the dimension of the tensors increases;
for this reason, the speedup of fast numerical max-convolution becomes
even more pronounced as the dimension increases. For a tensor of
dimension $d$ and width $k$ (\emph{i.e.}, where the index bounds of
every dimension are $\in \{0, 1, \ldots k-1\}$), the cost of naive
max-convolution will be in $O(k^{2d})$, whereas the cost of numerical
max-convolution is $O(k^d \log_2(k))$ (ignoring the
$\log_2(\log_{1+\tau^\frac{1}{4}}(k)) \leq 18$ multiplier), meaning
that there is an $O(\frac{k^d}{d\,\log_2(k)})$ speedup from the numerical
approach. Examples of such tensor problems include graph theory, where
adjacency matrix representation can be used to describe respective
distances between nodes in a network.

As a concrete example, the demonstration Python code computes
the max-convolution between two $256 \times 256$ matrices. The naive
method required $494$ seconds, but the numerical result with the
original piecewise method was computed in $3.18$ seconds (yielding a
maximum absolute error of $0.0173$ and a maximum relative error of
$0.0511$) and the numerical result with the null space projection
method was computed in $3.99$ seconds (using $p^*_{\max} = 512$,
which corresponds to a relative error of $<0.1\%$ in the top contour,
yielding a maximum absolute error of $0.0141$ and a maximum relative
error of $0.0227$) and in $3.05$ seconds (using $p^*_{\max} = 64$,
which corresponds to a relative error of $<2.5\%$ in the top contour,
yielding a maximum absolute error of $0.0667$ and a maximum relative
error of $0.067$). Not only does the speedup of the proposed methods
relative to naive max-convolution increase significantly as the
dimension of the tensor is increased, no other faster-than-naive
algorithms exist for max-convolution of matrices or tensors.

Multidimensional max-convolution can likewise be applied to hidden
Markov models with additive transitions over multidimensional
variables (\emph{e.g.}, allowing the latent variable to be a
two-dimensional joint distribution of American and German unemployment
with a two-dimensional joint transition probability).

\subsection{Max-Deconvolution}
The same $p^*$-norm approximation can also be applied to the problem
of max-deconvolution (\emph{i.e.}, solving $M = L *_{\max} R$ for $R$
when given $M$ and $L$). This can be accomplished by computing the
ratio of $FFT(M^{p^*})$ to $FFT(L^{p^*})$ (assuming $L$ has already
been properly zero-padded), and then computing the inverse FFT of the
result to approximate $R^{p^*}$; however, it should be noted that
deconvolution methods are typically less stable than the corresponding
convolution methods, computing a ratio is less stable than computing a
product (particularly when the denominator is close to zero).

\subsection{Amortized Argument for Low MSE of the Affine Piecewise Method}
Although the largest absolute error of the affine piecewise method is
the same as the largest absolute error of the original piecewise
method, the mean squared error (MSE) of the affine piecewise method
will be lower than the square of the worst-case absolute error.

To achieve the worst-case absolute error for a given contour the
affine correction must be negligible; therefore, there must be two
nearly vertical points on the scatter plot of $\| u^{(m_1)} \|_\infty$
vs. $\| u^{(m_1)} \|_{p^*}$, which are both extremes of the bounding
envelope from {\bf Figure~\ref{figure:approxExactLinearZoom}}. Thus,
there must exist two different indices $m_1$ and $m_2$ with vectors
where $\| u^{(m_1)} \|_{p^*} \approx \| u^{(m_1)} \|_\infty$ and where
\[\| u^{(m_2)} \|_{p^*} \approx \| u^{(m_2)} \|_\infty k_{m_2}^\frac{1}{p^*}\]
(creating two vertical points on the scatter plot, and forcing that
both cannot simultaneously be corrected by a single affine
mapping). In order to do this, it is required to have $u^{(m_1)}$
filled with a single nonzero value and for the remaining elements of
$u^{(m_1)}$ to equal zero. Conversely, $u^{(m_2)}$ must be filled
entirely with large, nonzero values (the largest values possible that
would still use the same contour $p^*$). Together, these two arguments
place strong constraints on the vectors $L'$ and $R'$ (and
transitively, also constrains the unscaled vectors $L$ and $R$): On
one hand, filling $u^{(m_1)}$ with $k_{m_1}-1$ zeros requires that
$k_{m_1}-1$ elements from either $L$ or $R$ must be zero (because at
least one factor must be zero to achieve a product of zero). On the
other hand, filling $u^{(m_2)}$ with all large-value nonzeros requires
that $k_{m_2}$ elements of \emph{both} $L$ and $R$ are
nonzero. Together, these requirements stipulate that both $k_{m_1}-1 +
k_{m_2} \leq k$, because entries of $L$ and $R$ cannot simultaneously
be zero and nonzero. Therefore, in order to have many such vertical
points, constrains the lengths of the $u^{(m_1)}, u^{(m_2)},
u^{(m_3)}, \ldots$ vectors corresponding to those points. While the
worst-case absolute error bound presumes that an individual vector
$u^{(m)}$ may have length $k$, this will not be possible for many
vectors corresponding to vertical points on the scatter plot. For this
reason, the MSE will be significantly lower than the square of the
worst-case absolute error, because making a high affine-corrected
absolute error on one index necessitates that the absolute errors at
another index cannot be the worst-case absolute error (if the sizes of
$L$ and $R$ are fixed).

\section{Availability}

Code for exact max-convolution and the fast numerical method (which
includes both $\| \cdot \|_{p^*}$ and null space projection methods)
is implemented in Python and available at
\url{https://bitbucket.org/orserang/fast-numerical-max-convolution}. All
included code works for {\tt numpy} arrays of any dimension,
\emph{i.e.}  tensors).

\section{Acknowledgments}

We would like to thank Mattias Fr{\aa}nberg, Knut Reinert, and Oliver
Kohlbacher for the interesting discussions and
suggestions. J.P. acknowledges funding from BMBF (Center for
Integrative Bioinformatics, grant no. 031A367). O.S. acknowledges
generous start-up funds from Freie Universit\"{a}t Berlin and the
Leibniz-Institute for Freshwater Ecology and Inland Fisheries.


\end{document}